\documentclass[journal]{IEEEtran}
\usepackage{bm}
\usepackage{amsfonts}
\usepackage{amsmath}

\usepackage{graphicx}  
\usepackage{cite}
\usepackage{epstopdf}
\usepackage{multirow}
\usepackage{booktabs}

\begin{document}

\title{Dimensionality Increment of PMU Data for Anomaly Detection in Low Observability Power Systems}

\author{ Xin Shi$^1$,~\IEEEmembership{Student Member,~IEEE}, Robert Qiu$^{1,2}$,~\IEEEmembership{Fellow,~IEEE}
\thanks{This work was partly supported by National Key R \& D Program of China under Grant 2018YFF0214705, NSF of China under Grant 61571296 and (US) NSF under Grant CNS-1619250.

$^1$ Department of Electrical Engineering,Center for Big Data and Artificial Intelligence, State Energy Smart Grid Research and Development Center, Shanghai Jiaotong University, Shanghai 200240, China.(e-mail: dugushixin@sjtu.edu.cn; rcqiu@sjtu.edu.cn.)

$^2$ Department of Electrical and Computer Engineering,Tennessee Technological University, Cookeville, TN 38505, USA. (e-mail:rqiu@tntech.edu)

}
}

\maketitle

\begin{abstract}
Anomaly detection is an important task in power systems. To make better use of the phasor measurement unit (PMU) data collected from a low observability power system for anomaly detection, a data dimensionality increment algorithm is proposed in this paper. First, a low-dimensional spatio-temporal data matrix is formulated by using the synchrophasor measurements collected from a limited number of PMUs in a power system. Then, a data dimensionality increment algorithm based on random tensor theory (RTT) is proposed for anomaly detection. The proposed algorithm can help improve the sensitivity of random matrix theory (RMT) based and machine learning (ML) based anomaly detection approaches, and it is able to accelerate the convergence rate of model training in the ML based anomaly detection approach. Case studies on the IEEE 118-bus test system validate the effectiveness of the proposed algorithm.
\end{abstract}

\IEEEpeerreviewmaketitle

\section{Introduction}
\label{section: Introduction}

\IEEEPARstart{T}{his} paper is driven by the need of anomaly detection to make better use of the PMU data from low observability power systems. Anomaly detection is a fundamental task in power systems, which can help realize the situation awareness of the systems and offer support on the safety analysis and control decision. In recent years, there have been increasing deployments of PMUs in power systems, which constitute the wide area measurement system (WAMS) \cite{zima2005design}. Compared with traditional supervisory control and data acquisition (SCADA) system, WAMS can provide synchrophasor measurements with higher sampling rates, which makes it possible for real-time anomaly detection.

The synchophasor data collected from WAMS contain rich information on the operating states of power systems. By leveraging the data, a variety of data-driven approaches are developed for anomaly detection. The anomaly detection approaches can be roughly categorized into three classes: 1) statistical approaches, 2) signal processing approaches, and 3) artificial intelligence approaches. The statistical approaches often use simple calculated indexes, such as the maximum (or minimum), mean, variance, etc \cite{guralnik1999event,yang1998study,allen2012algorithm}. The signal processing approaches, frequently used in recent years, include fourier and wavelet transform analysis \cite{santoso2000characterization,kim2017wavelet}, principal component analysis (PCA) \cite{xie2014dimensionality}, RMT \cite{he2017big,shi2018saptio,shi2018incipient,shi2019early}, etc. The artificial intelligence approaches, especially the machine learning approaches, include one-class support vector machine (OSVM) \cite{ma2003time,ERFANI2016121}, stacked auto-encoder (SAE) networks \cite{sakurada2014anomaly,liu2018anomaly}, long short term memory (LSTM) networks \cite{malhotra2015long,malhotra2016lstm}, etc. These approaches have been proved to be powerful in anomaly detection.

In practice, it is neither economical nor necessary to install a PMU at every bus of a power system, because the PMU is costly and the voltage/current (phasor) of the incident buses to a PMU installed bus can be calculated through branch parameters \cite{Nuqui2005Phasor,Aminifar2010Contingency,Roy2012An}. Therefore, only a limited number of PMUs are installed for a power system. For example, reference \cite{Baldwin1993Power} reports that PMUs need to be installed at 1/5 to 1/3 of the number of system buses for the system to be observable. For a low observability power system, the synchrophasor data collected from a limited number of PMUs are often low-dimensional. Here, each PMU is considered as an dimension and the dimensionality of the data is equal to the number of PMUs. Due to the low data dimensionality, performances of many powerful anomaly detection approaches are limited. For example, for the RMT based approaches, infinite or high data dimensionality is required for the asymptotic theorems in theory, and the low dimensionality will cause inaccurate analysis results; for the ML based approaches, the low dimensional data are not sufficient to train more powerful prediction models.

In view of the limitations of low synchrophasor data dimensionality for current anomaly detection approaches, this paper presents a fundamental data dimensionality increment algorithm based on random tensor theory. It is a new breakthrough in probability and statistics for proving an alternative way to study independence, which makes it possible for the analysis of large dimensional data by using many advanced mathematical tools, such as asymptotic theorems in RMT, concentration inequality, free probability, etc. The main contributions of this paper are summarized as follows: 1) The random tensor theory, including tensor product of random vectors, tensor version of sample covariance matrix and linear eigenvalue statistics (LES), is introduced and remarked. 2) For the synchrophasor data from a low observability power system, a data dimensionality increment algorithm based on the RTT is proposed. 3) It is experimentally justified that the proposed algorithm can help improve the anomaly detection sensitivity of RMT based and ML based approaches. 4) It is experimentally showed that the proposed algorithm is able to accelerate the convergence rate of model training in the ML based anomaly detection approach.

The rest of this paper is organized as follows. In Section \ref{section: analysis}, the RTT is introduced and remarked. In Section \ref{section: anomaly detection}, a spatio-temporal data matrix is formulated by arranging low-dimensional synchrophasor measurements in chronological order, and a data dimensionality increment algorithm based on the RTT is proposed. The steps of the proposed increasing data dimensionality for RMT based and ML based anomaly detection approaches are given. Section \ref{section: case} validates the effectiveness of the proposed algorithm on the IEEE 118-bus test system. Conclusions are presented in Section \ref{section: conclusion}.

\section{Random Tensor Theory}
\label{section: analysis}
In this section, the RTT is introduced and remarked. First, the tensor product of random vectors is defined. Based on this, tensor version of sample covariance matrix is introduced and the steps for constructing it is given. Then the comparison of the empirical spectral distributions (ESDs) of traditional covariance matrix and the tensor version with their theoretical limits is made and analyzed. The LES for the tensor version sample covariance matrix is defined.
\subsection{Tensor Product of Random Vectors}
\label{subsection: tensor_product_vector}
Let ${\bf a}=[a_1,\cdots,a_i]^H\in{\mathbb C}^i$ and ${\bf b}=[b_1,\cdots,b_j]^H\in{\mathbb C}^j$, the tensor (Kronecker) product of vector $\bf a$ and $\bf b$ is defined as
\begin{equation}
\label{Eq:tensor_product}
\begin{aligned}
  {\bf a}\otimes {\bf b} = [a_1b_1,\cdots,a_1b_j,\cdots\cdots,a_ib_1,\cdots,a_ib_j]^H
\end{aligned},
\end{equation}
where $\otimes$ denotes the tensor product operation and ${\bf a}\otimes {\bf b}$ is a vector of size $ij$.

Assume a random vector ${\bf x}^{(0)}=[x_1,\cdots,x_n]\in{\mathbb C}^n$, we can naturally construct a random vector $\bf x$ by using the tensor products of $k$ independent identically distributed (i.i.d.) copies of ${\bf x}^{(0)}$, namely
\begin{equation}
\label{Eq:tensor_product_vectors}
\begin{aligned}
  {\bf x}={\bf x}^{(1)}\otimes\cdots\otimes{\bf x}^{(k)}
\end{aligned},
\end{equation}
where ${\bf x}^{(1)},\cdots,{\bf x}^{(k)}\in{\mathbb C}^n$ are i.i.d. copies of the random vector ${\bf x}^{(0)}$, and the new random vector ${\bf x}\in{\mathbb C}^{n^k}$ lies in the $n^k$ dimensional normed space.
Equation (\ref{Eq:tensor_product_vectors}) maps a random vector of size $n\times k$ to higher dimensionality $n^k$ without any additional conditions, which makes it possible for the use of many advanced tools that require high dimensionality as the prerequisites in statistics and machine learning.
\subsection{Tensor Version of Sample Covariance Matrix}
\label{subsection: tensor_product_covariance}
In Section \ref{subsection: tensor_product_vector}, the dimensionality of a random vector can be greatly increased by using the tensor product operation. The traditional asymptotic approaches \cite{anderson1962} in statistics are developed for low dimensional problems, which will cause accumulating errors of estimates of a large number of parameters if directly extended to high dimensional problems \cite{serdobol1999theory}. In recent years, it has been well studied for the scenario that both the dimensionality and sample size go to infinity at the same rate \cite{ambainis2012random,lytova2017central,Qiu2017}.

For the dimensionality increased random vector ${\bf x}\in{\mathbb C}^{n^k}$ in Section \ref{subsection: tensor_product_vector}, consider $n^k\times n^k$ random matrices of the form
\begin{equation}
\label{Eq:tensor_covariance}
\begin{aligned}
  {{\bf{\mathcal{M}}}_{n,N,k}}\left( {\bf{x}} \right) = \sum\limits_{\alpha = 1}^N {{\tau _\alpha}{{{\bf{x}}}_\alpha}{\bf{x}}_\alpha^H},\quad {\bf x}_\alpha={\bf x}_{\alpha}^{(1)}\otimes\cdots\otimes{\bf x}_{\alpha}^{(k)}
\end{aligned},
\end{equation}
where $\tau_\alpha(\alpha=1,2,\cdots,N)$ are real numbers, and ${{\bf{\mathcal{M}}}_{n,N,k}}({\bf{x}})$ is the tensor version of sample covariance matrix. For every fixed $k\ge 1$, as $N\to\infty, n\to\infty,$ but $\frac{N}{n^k} \to c\in (0,\infty)$, the ESD of ${{\bf{\mathcal{M}}}_{n,N,k}}({\bf{x}})$ converges to a non-random measure. Steps for constructing the tensor version of sample covariance matrix is shown in Algorithm 1.
\begin{table}[htbp]
\label{Tab: steps_approach}
\centering
\begin{tabular}{p{8.4cm}}   
\toprule[1.0pt]
\textbf {Algorithm 1:} Steps for constructing the tensor version of sample covariance matrix\\
\hline
1. \textbf{For} the index $\alpha=1,2,\cdots,N$: \\
2.\quad Generate $k$ i.i.d. copies ${\bf x}_{\alpha}^{(1)},\cdots,{\bf x}_{\alpha}^{(k)}$ of the normalized \\
\qquad random vector ${\bf x}_{\alpha}^{(0)}\in{\mathbb C}^{n}$. \\
3.\quad Construct a high dimensional random vector ${\bf x}_\alpha$ using the tensor \\
\qquad product of the generated $k$ random vectors, i.e., \\
\qquad ${\bf x}_\alpha={\bf x}_{\alpha}^{(1)}\otimes\cdots\otimes{\bf x}_{\alpha}^{(k)}$. \\
4.\quad Form the rank-one random matrix ${\bf x}_\alpha{\bf x}_\alpha^H$. \\
\quad \textbf{End For} \\
5. Obtain the sum of those weighted random matrices, i.e., ${{\bf{\mathcal{M}}}_{n,N,k}}({\bf{x}})$. \\
\hline
\end{tabular}
\end{table}

\begin{figure}[htb]
\centering
\begin{minipage}{4.1cm}
\centerline{
\includegraphics[width=1.75in]{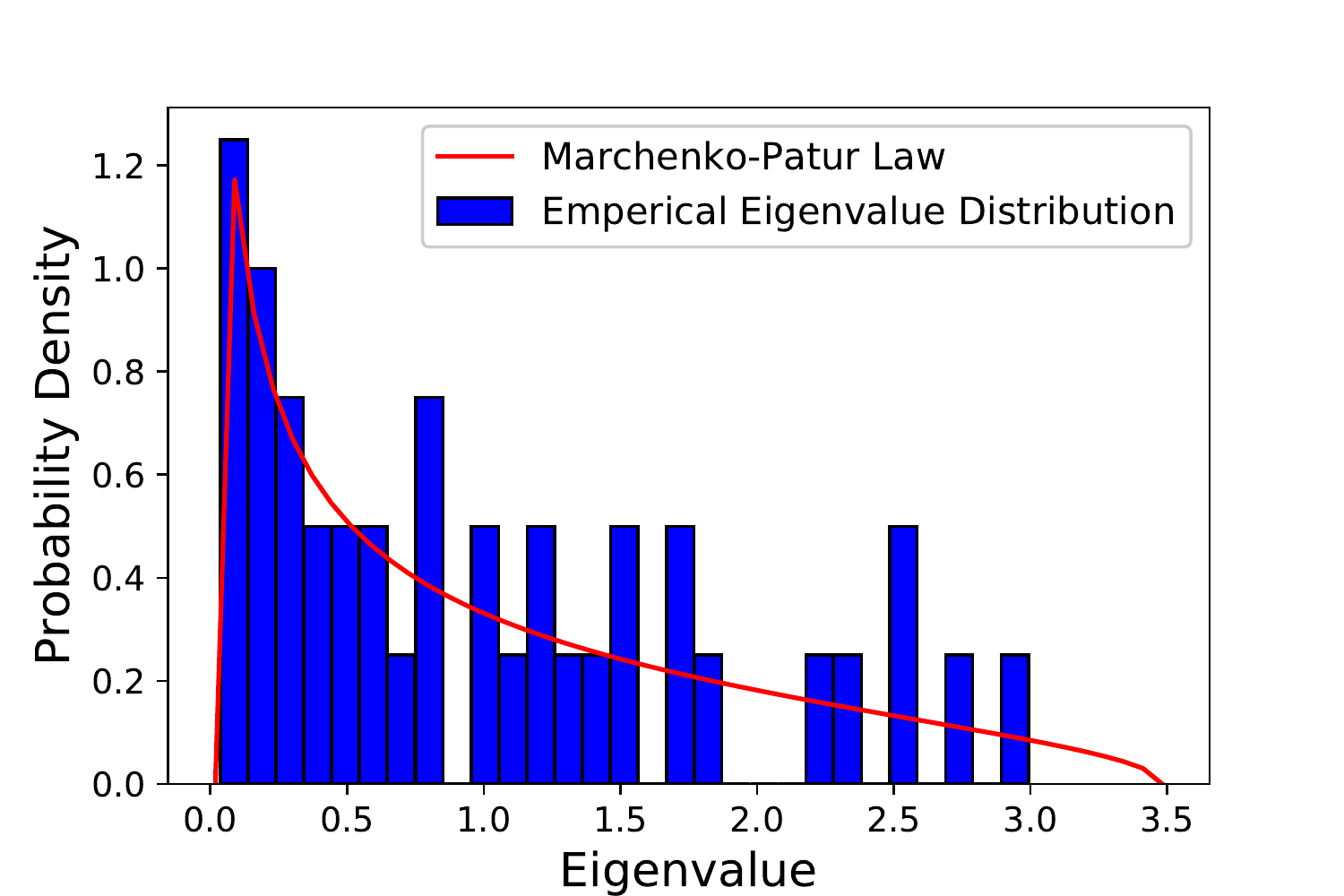}
}
\parbox{5cm}{\small \hspace{0.5cm}(a1) Dimensionality: $54$ }
\end{minipage}
\hspace{0.1cm}
\begin{minipage}{4.1cm}
\centerline{
\includegraphics[width=1.75in]{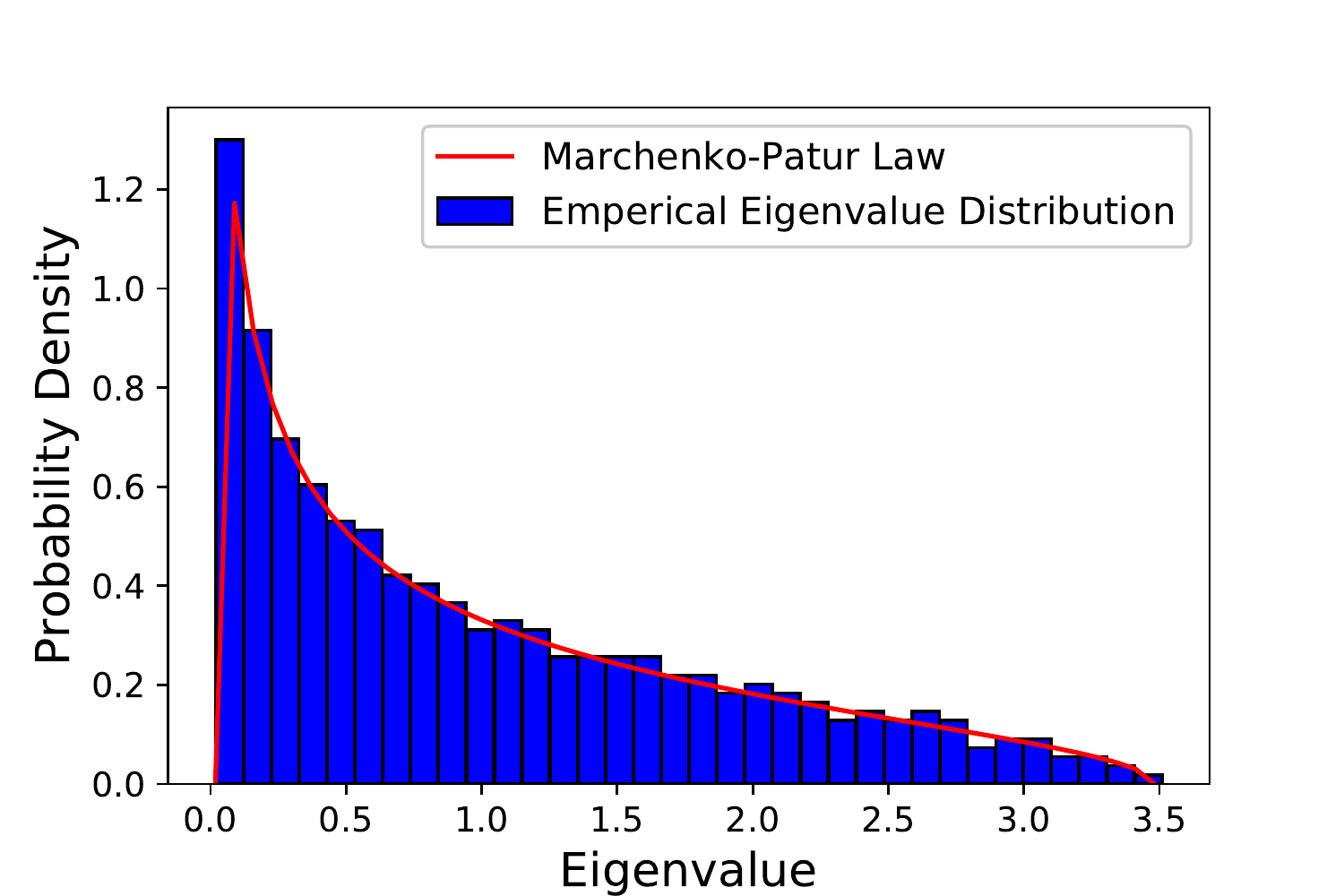}
}
\parbox{5cm}{\small \hspace{0.5cm}(a2) Dimensionality: $729$ }
\end{minipage}
\hspace{0.1cm}
\begin{minipage}{4.1cm}
\centerline{
\includegraphics[width=1.75in]{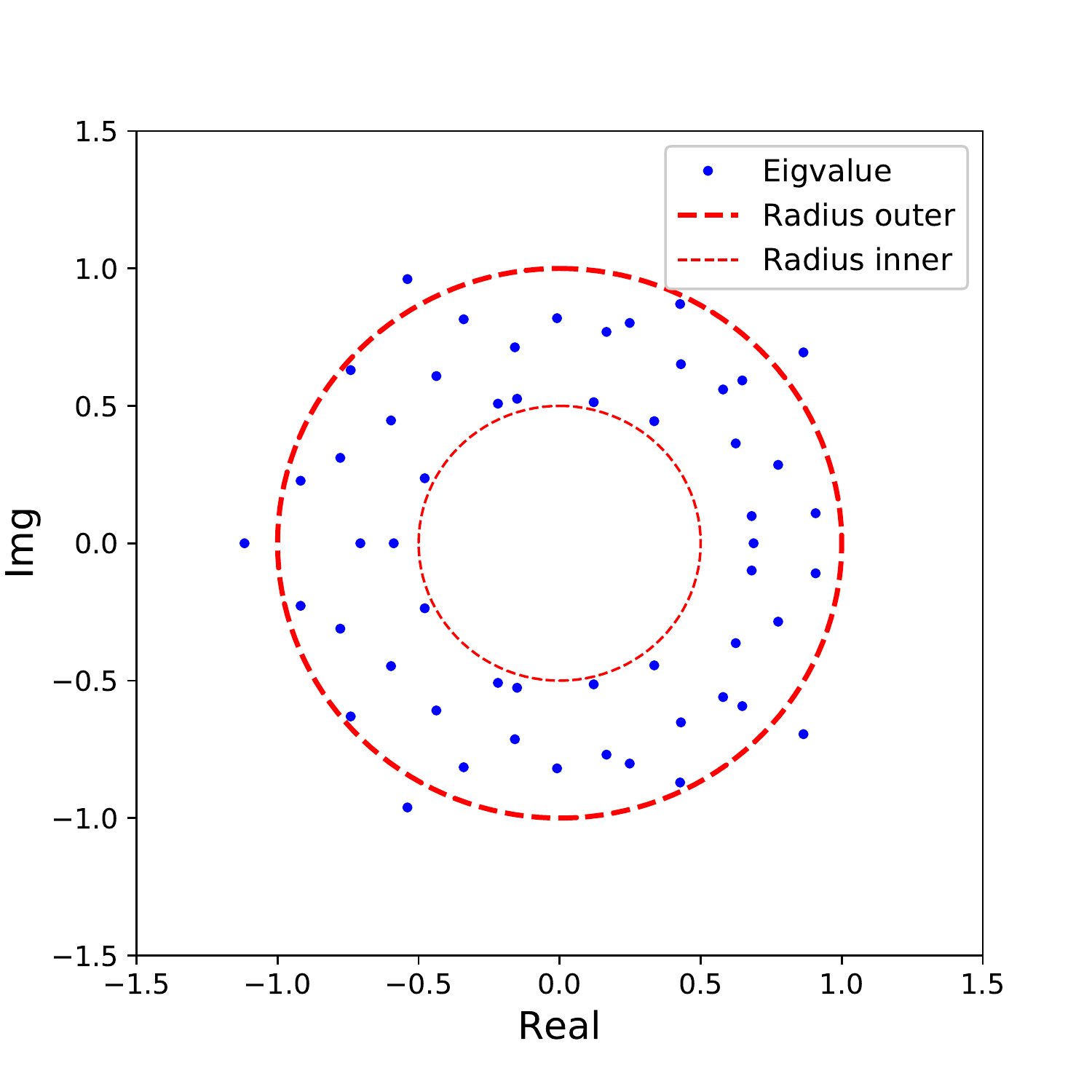}
}
\parbox{5cm}{\small \hspace{0.5cm}(b1) Dimensionality: $54$ }
\end{minipage}
\hspace{0.1cm}
\begin{minipage}{4.1cm}
\centerline{
\includegraphics[width=1.75in]{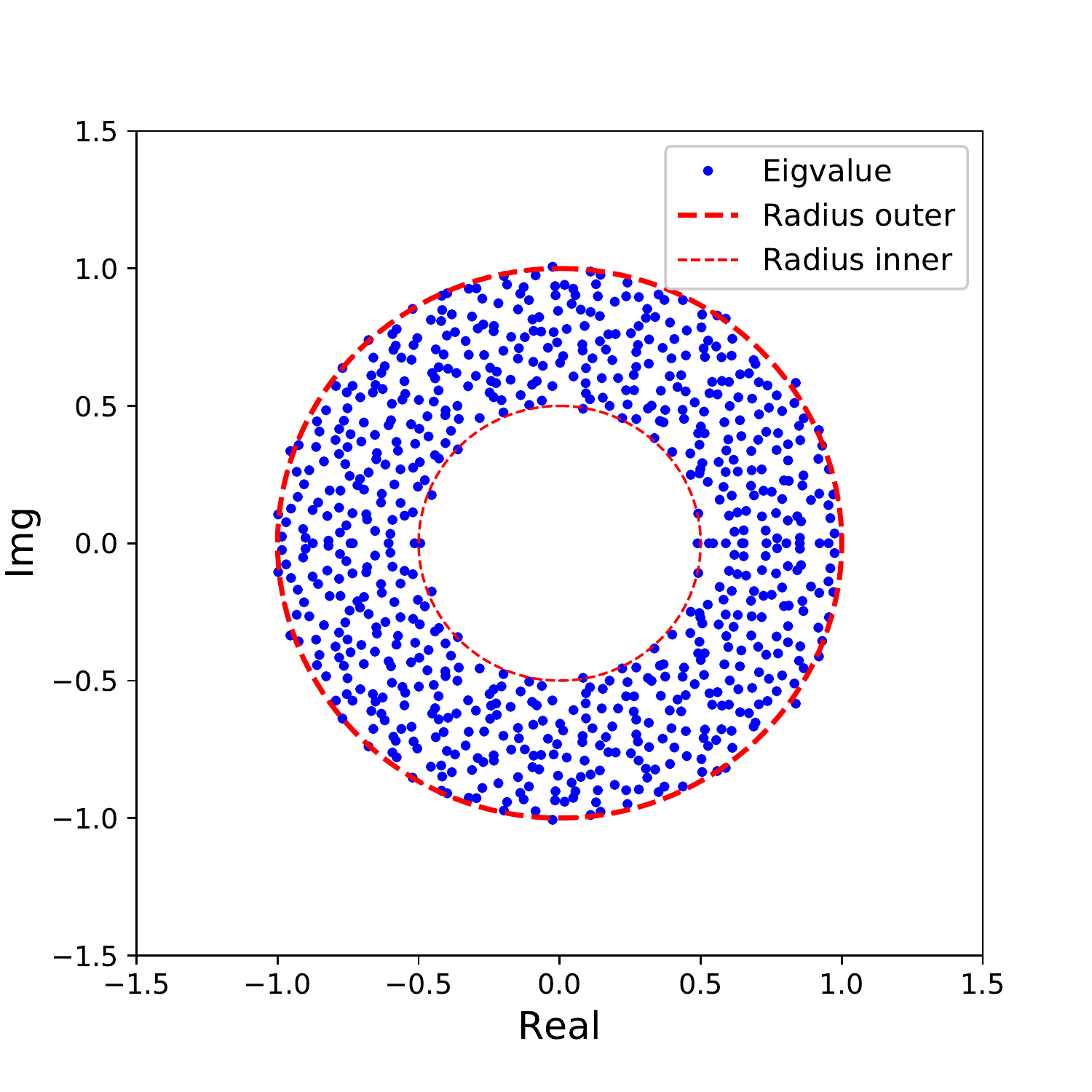}
}
\parbox{5cm}{\small \hspace{0.5cm}(b2) Dimensionality: $729$ }
\end{minipage}
\caption{The ESDs of traditional covariance matrix and the tensor version, and their comparisons with the theoretical M-P law and Ring law. The dimensionality of data vector ${\bf x}_\alpha$ is increased from $n\times k=27\times 2=54$ to $n^k=27^2=729$ through the tensor product operation, the radio $c$ is $0.75$, and $\tau _\alpha$ is $1$.}
\label{fig:law_tensor}
\end{figure}
Comparing the ESDs of traditional covariance matrix and the tensor version with their theoretical limits is insightful. Figure \ref{fig:law_tensor} shows the ESD of traditional covariance random matrix and the tensor version and their comparisons with the theoretical Marchenko-Pastur (M-P) law \cite{marvcenko1967distribution} and Ring law \cite{guionnet2009single,ipsen2014weak}. It can be observed that the ESD of the traditional covariance random matrix does not fit the theoretical limits well for the reason of low dimensionality. In contrast, the ESD of the tensor version covariance random matrix converges almost surely to the theoretical limits.
\subsection{Linear Eigenvalue Statistics}
\label{subsection: les}
The LES is a high-dimensional statistic for the eigenvalues of ${{\bf{\mathcal{M}}}_{n,N,k}}$ in equation (\ref{Eq:tensor_covariance}), which is defined as
\begin{equation}
\label{Eq:les_tensor}
\begin{aligned}
  \mathcal{N}_n[\varphi]=\sum\limits_{i=1}^{n^k}\varphi (\lambda_i^{(n)})
\end{aligned},
\end{equation}
where $\lambda_i$ are the eigenvalues, and $\varphi(\cdot)$ is a test function which maps the eigenvalues into high dimensional space. The commonly used test functions include Chebyshev Polynomial ($\varphi (\lambda)={a_0}{\lambda}^n+{a_1}{\lambda}^{n-1}+\cdots+{a_n}$, where $a_i (i=1,2,\cdots,n)$ are coefficients), information entropy ($\varphi (\lambda)=-{\lambda}log{\lambda}$), likelihood radio function ($\varphi (\lambda)={\lambda}-log{\lambda}-1$), etc \cite{Qiu2013Cognitive}. For a bounded continuous function $\varphi$, we have in probability
\begin{equation}
\label{Eq:les_tensor_limit}
\begin{aligned}
  \lim _{n \rightarrow \infty} n^{-k} \mathcal{N}_{n}[\varphi]=\int \varphi(\lambda) d N(\lambda)
\end{aligned}.
\end{equation}
Since the limit is non-random, the central limit theorem (CLT) has been proved for $\mathcal{N}_{n}[\varphi]$ of ${{\bf{\mathcal{M}}}_{n,N,1}}$ and ${{\bf{\mathcal{M}}}_{n,N,2}}$ in \cite{guedon2014central} and \cite{lytova2017central}.

In the complex plane of eigenvalues, the mean spectral radius (MSR), which can be regarded as a special form of LES, is a statistic for the eigenvalues of the singular value equivalent \cite{ipsen2014weak} matrix of ${{\bf{\mathcal{M}}}_{n,N,k}}$. It is the mean distribution radius of eigenvalues, which can be defined as
\begin{equation}
\label{Eq:msr_tensor}
\begin{aligned}
  {\kappa _{MSR}} = \frac{1}{n^k}\sum\limits_{i = 1}^{n^k} {\left| {{\lambda _i^{(n)}}} \right|}
\end{aligned},
\end{equation}
where $| \lambda _i^{(n)} |$ is the radius of the eigenvalue $\lambda _i^{(n)}$ on the complex plane.

\section{Anomaly Detection Using PMU Data}
\label{section: anomaly detection}
Based on the RTT in Section \ref{section: analysis}, a PMU data dimensionality increment algorithm is proposed. First, a spatio-temporal data matrix is formulated by arranging the synchrophasor measurements collected from a low observability power system in chronological order. Then, details on the RMT based and ML based anomaly detection approaches are presented. Finally, the steps of the proposed increasing data dimensionality for RMT based and ML based anomaly detection approaches are given.
\subsection{Formulation of PMU Data as a Spatio-Temporal Matrix}
\label{subsection: formulation}
\begin{figure}[!t]
\centerline{
\includegraphics[width=3.0in]{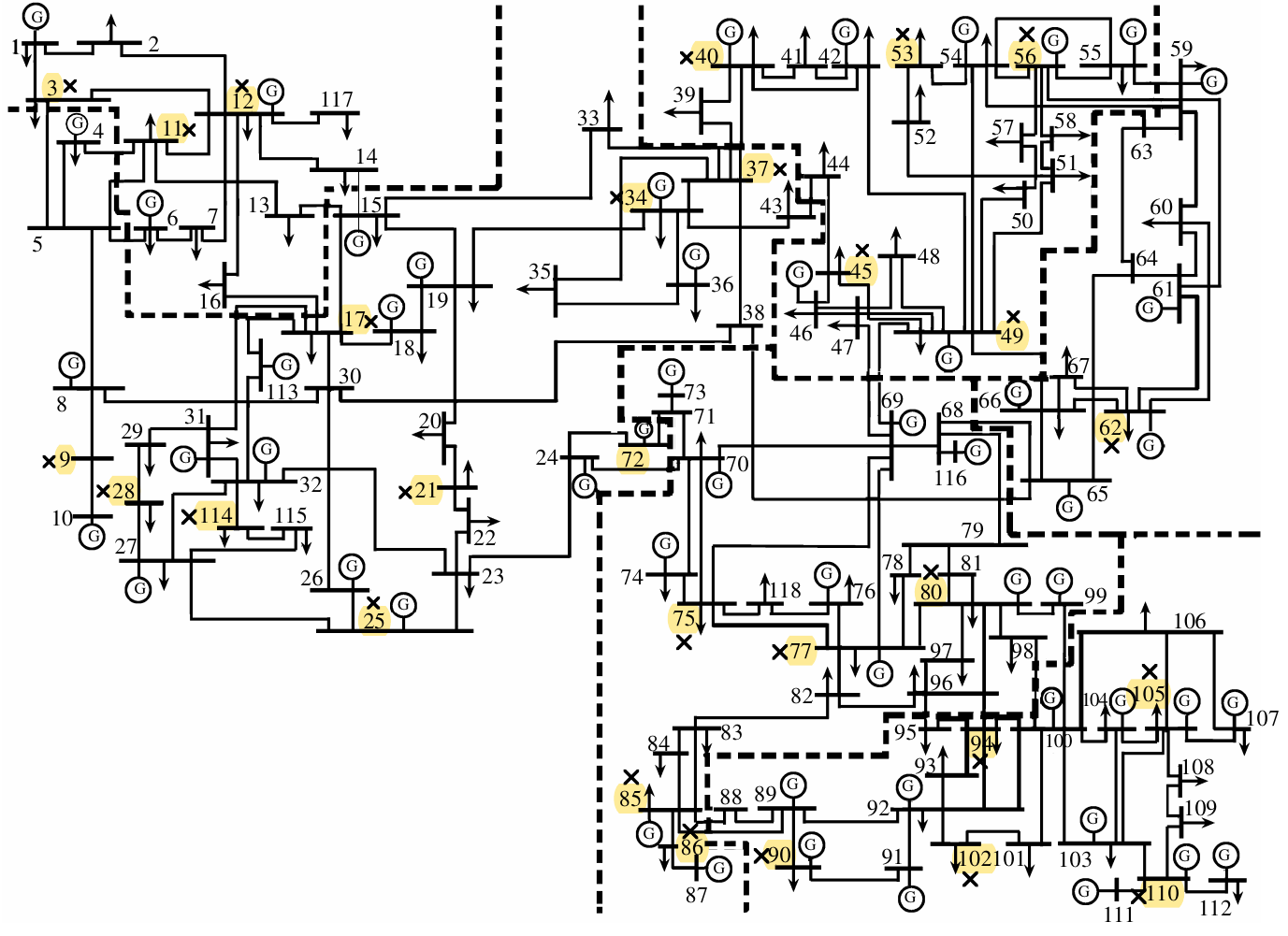}
}
\caption{PMU placement location for the IEEE 118-bus test system. The PMU placement buses have been marked with a cross.}
\label{fig:pmu_placement}
\end{figure}

Assume there are $P$-dimensional measurement variables (such as $P-$dimensional voltage measurements from $P$ PMUs installed in a power system) $(d_1,d_2,...,d_P)\in \mathbb{C}^{1\times P}$. At the sampling time $t_j$, the $P-$dimensional measurements can be formulated as a column vector ${\bf d}(t_j)=(d_1,d_2,...,d_P)^H$. For a series of time $N$, a spatio-temporal data matrix ${\bf D}\in \mathbb{C}^{P\times N}$ is formulated by arranging these vectors $\bf d$ in chronological order. To be mentioned is that, by stacking the $P$ measurements in a series of time $N$ together, the spatio-temporal data matrix contains the most information on the operating states of the system. For example, 28 PMUs must be installed in an IEEE 118-bus test system in order to make the system observable \cite{Aminifar2010Contingency,Roy2012An}, which is shown in Figure \ref{fig:pmu_placement}. Thus, for a series of  $1000$ sampling times, a $28\times 1000$ matrix of voltage measurements can be formulated.

\subsection{Dimensionality Increment of PMU Data for Anomaly Detection}
\label{subsection: anomaly_detection}
For the formulated spatio-temporal data matrix ${\bf D}=[{{\bf d}_1},{{\bf d}_2},\cdots,{{\bf d}_N}]\in \mathbb{C}^{P\times N}$ in Section \ref{subsection: formulation}, where ${\bf d}_i\;(i=1,\cdots,N)$ are the column vectors. Let $P=kn$ ($k,n\in\mathbb{N}$), thus the column vector ${\bf d}_i$ can be partitioned into $k$ vectors ${\bf{d}}_i^{(l)}$ ($l=1,2,\cdots,k$), each of size $n$. We first normalize those partitioned vectors by
\begin{equation}
\label{Eq:normalize_vector}
\begin{aligned}
  {\bf\tilde{d}}_i^{(l)} = \frac{{\bf{d}}_i^{(l)}}{||{\bf{d}}_i^{(l)}||}
\end{aligned},
\end{equation}
where $||\cdot||$ denotes the Euclidean norm.

Then, for each column vector, we can construct a higher dimensional vector by using the tensor products of those normalized vectors in the form
\begin{equation}
\label{Eq:tensor_pmu_data}
\begin{aligned}
  {{\bf\tilde{d}}_i} = {\bf\tilde{d}}_i^{(1)} \otimes  \cdots  \otimes {\bf\tilde{d}}_i^{(k)} \in {\left( {{\mathbb{C}^n}} \right)^{ \otimes k}}
\end{aligned},
\end{equation}
where ${\bf\tilde d}_i$ is the new constructed vector which lies in a $n^k$ dimensional normed space. Thus the dimensionality increased spatio-temporal data matrix ${\bf\tilde D}=[{{\bf\tilde d}_1},{{\bf\tilde d}_2},\cdots,{{\bf\tilde d}_N}]\in \mathbb{C}^{n^k\times N}$ is obtained.

In the RMT based anomaly detection approach, a $n^k\times N'$ ($N'<N$) window $\bf X$ is moved on ${\bf\tilde D}$ at continuous sampling times and the last sampling time is considered as the current time. For example, at the sampling time $t_i$, the obtained data window ${\bf X}(t_i)$ is formulated as
\begin{equation}
\label{Eq:data_window}
\begin{aligned}
  {\bf{X}}(t_i) = \left[ {{\bf\tilde{d}}(t_{i - N' + 1}),{\bf\tilde{d}}(t_{i - N' + 2}), \cdots ,{\bf\tilde{d}}(t_i)} \right]
\end{aligned},
\end{equation}
where ${\bf\tilde d}(t_j)={({\tilde d_1,\tilde d_2,\cdots, \tilde d_{n^k}})}^H$ ($t_{i-N'+1}\le t_j \le t_i$). Then the tensor version sample covariance matrix of ${\bf X}(t_i)$ is calculated as
\begin{equation}
\label{Eq:tensor_covariance_pmu}
\begin{aligned}
  {{\bf{\mathcal{M}}}_{n,N',k}}\left( {{\bf X}(t_i)} \right) = \sum\limits_{j = i-N'+1}^{i} {{\tau (t_j) }{{{\bf\tilde{d}}}(t_j)}({\bf\tilde{d}}(t_j))^H}
\end{aligned},
\end{equation}
where $\tau (t_j)$ are real numbers. Then the eigenvalues of ${{\bf{\mathcal{M}}}_{n,N',k}}\left( {{\bf X}(t_i)} \right)$ in the real and complex plane can be calculated. Thus, $\mathcal{N}_n[\varphi]$ in Equation (\ref{Eq:les_tensor}) and $\kappa_{MSR}$ in Equation (\ref{Eq:msr_tensor}) are generated for each sampling time with continuously moving windows, which enables us to track the data behavior effectively.

In the ML based anomaly detection approach, we first train a one-class prediction model by using the normal data set, and then compute the prediction error for the target data to detect anomalies. Considering anomaly rarely happens in power systems and anomalous data are difficult to collect, the unsupervised learning way is used here. Let ${\bf S}_{train}\subset {\bf\tilde D}$ be the normal data set, the prediction model is trained by using the sequence in ${\bf S}_{train}$, where each sampling data ${\bf\tilde d}_i=[{\tilde d_{i,1}},{\tilde d_{i,2}},\cdots,{\tilde d_{i,n^k}}]^H\in {{\bf S}_{train}}$ is used as the input of the prediction model and the label is itself. Let ${\bf S}_{test}$ be the remaining data sequence of ${\bf\tilde D}$, the learned prediction model is used to compute an error vector ${\bf e}_j=[{e_{j,1}},{e_{j,2}},\cdots,{e_{j,n^k}}]^H$ for each sampling data ${\bf\tilde d}_j=[{\tilde d_{j,1}},{\tilde d_{j,2}},\cdots,{\tilde d_{j,n^k}}]^H\in {{\bf S}_{test}}$, where $e_{j,l}$ ($l=1,2,\cdots,n^k$) is the difference between ${\tilde d_{j,l}}$ and its value as predicted. The root mean squared error (RMSE) for the sampling data ${\bf\tilde d}_j$ is calculated as
\begin{equation}
\label{Eq:rmse}
\begin{aligned}
  RMSE({\bf\tilde d}_j)=\sqrt{\frac{1}{n^k}\sum\limits_{l=1}^{n^k} e_{j,l}^2}
\end{aligned},
\end{equation}
which is used as the anomaly indicator.

Based on the research above, steps of the proposed increasing PMU data dimensionality for RMT based and machine learning based anomaly detection approaches are summarized as in Algorithm 2.
\begin{table}[htbp]
\label{Tab: steps_approach}
\centering
\begin{tabular}{p{8.4cm}}   
\toprule[1.0pt]
\textbf {Algorithm 2:} Steps of dimensionality increment of PMU data for anomaly detection\\
\hline
1: A spatio-temporal data matrix ${\bf D}\in{\mathbb{C}^{P\times N}}$ is formulated by arranging \\
\quad $P$ synchrophasor measurements in a series of time $N$ in chronological \\
\quad order.  \\
2: For each column vector ${\bf d}_i\;(i=1,\cdots,N)$ of $\bf D$: \\
  \quad 2a) Partition ${\bf d}_i$ into $k$ vectors ${\bf{d}}_i^{(l)}$ ($l=1,2,\cdots,k$), each of size $n$; \\
  \quad 2b) Normalize ${\bf{d}}_i^{(l)}$ by Equation (\ref{Eq:normalize_vector}); \\
  \quad 2c) Construct a higher dimensional vector ${\bf\tilde d}_i$ through the tensor \\
  \quad\quad\; products of ${\bf{d}}_i^{(l)}$, i.e., Equation (\ref{Eq:tensor_pmu_data}). \\
3: Obtain the dimensionality increased data matrix ${\bf\tilde D}\in{\mathbb{C}^{n^k\times N}}$. \\
4: In the RMT based anomaly detection approach: \\
  \quad 4a) Move a $n^k\times N'$ ($N'<N$) window on ${\bf\tilde D}$ at continuous sampling \\
  \quad\quad\; times; \\
  \quad 4b) Calculate the tensor version sample covariance matrix ${{\bf{\mathcal{M}}}_{n,N',k}}$ \\
  \quad\quad\; of the data window at each sampling time; \\
  \quad 4c) Calculate the eigenvalues of ${{\bf{\mathcal{M}}}_{n,N',k}}$ in the real and complex \\
  \quad\quad\; plane for each sampling time; \\
  \quad 4d) Generate $\mathcal{N}_n[\varphi]-t$ and $\kappa_{MSR}-t$ curves through Equation (\ref{Eq:les_tensor}) \\
  \quad\quad\; and (\ref{Eq:msr_tensor}), respectively. \\
5: In the ML based anomaly detection approach: \\
  \quad 5a) Train a one-class prediction model using the normal data set \\
  \quad\quad\; ${\bf S}_{train}\subset{\bf\tilde D}$; \\
  \quad 5b) Compute the root mean squared error for each sampling of the \\
  \quad\quad\; target data ${\bf S}_{test}\subset{\bf\tilde D}$ to detect anomalies. \\
\hline
\end{tabular}
\end{table}

\section{Case Studies}
\label{section: case}
In this section, the effectiveness of the proposed algorithm is validated with the synthetic data collected from a low observability IEEE 118-bus test system \cite{5491276}. Detailed information about the IEEE 118-bus test system can be found in case118.m in Matpower6.0 package \cite{zimmerman2016matpower}. The details of PMU placement location for the IEEE 118-bus test system is shown in Figure \ref{fig:pmu_placement}. In the simulations, two types of anomalies were set: 1) the open circuit caused by the increase of impedance, and 2) the load change. Meanwhile, a little white noise was introduced to represent random fluctuations. For the generated synthetic data $\bf D$, a little colored noise $\bf E$ (i.e., $E_{i,t}=b*E_{i,t-1}+\varepsilon_{i,t}$, where $b$ is the correlation coefficient and $\varepsilon_{i,t}\sim N(0,1-b^2)$ so that the variance of $E_t$ is $1$.) was introduced to represent measuring errors. The scale of the colored noise is calculated by $m=\sqrt{\frac{var (\bf D)}{var {(\bf E)}*SNR}}$, where $var (\cdot)$ denotes the variance operation, and $SNR$ is the signal-noise-ratio.
\subsection{Case Studies on the Detection of Open Circuit}
\label{subsection: case_A}
\begin{figure}[!t]
\centerline{
\includegraphics[width=2.5in]{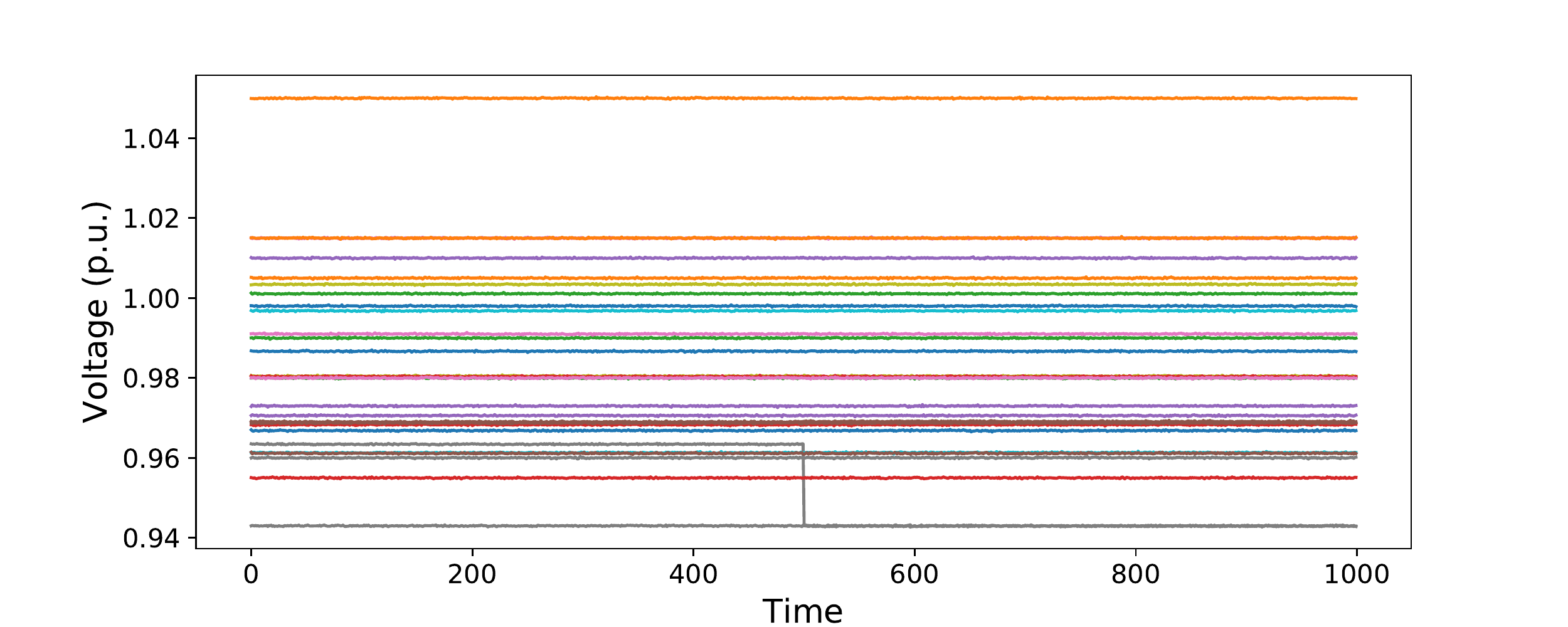}
}
\caption{The synthetic data collected from $28$ PMUs installed in the IEEE 118-bus test system in Case A. The open circuit anomaly was set at $t_s=501$.}
\label{fig:case1_data_org}
\end{figure}

In this section, the open circuit anomaly was set by a sudden increase of impedance from bus 29 to 31 in the IEEE 118-bus test system, which was shown in Table \ref{Tab: Case1}. The synthetic data collected from $28$ PMUs installed in the system was shown in Figure \ref{fig:case1_data_org}. It contained $28$ voltage measurement variables with sampling $1000$ times.
\begin{table}[!t]
\caption{The Open Circuit Anomaly From Bus $29$ to $31$ in Case A.}
\label{Tab: Case1}
\centering
\begin{tabular}{cclc}   
\toprule[1.0pt]
\textbf {fBus} & \textbf {tBus} & \textbf{Sampling Time}& \textbf{Impedance(p.u.)}\\
\hline
\multirow{2}*{29} & \multirow{2}*{31} & $t_s=1\sim 500$ & 0.02 \\
~ & ~ & $t_s=501\sim 1000$ & 20 \\
\hline
Others & Others & $t_s=1\sim 1000$ & Unchanged \\
\hline
\end{tabular}
\end{table}

1) RMT Based Anomaly Detection Approach: The effectiveness of the proposed increasing data dimensionality for RMT based anomaly detection approach is validated in this case. By using the proposed dimensionality increment approach, the dimensionality of the synthetic data was increased from $28=14\times 2$ to $196=14^2$. In the experiments, the RMT based anomaly detection approach was used for analyzing both the original data and the dimensionality increment one. The size of the moving window was set as $P\times 200$, where $P$ was $28/196$. The signal-noise-ratio $SNR$ was set to be $1000$ and the test function in Equation (\ref{Eq:les_tensor}) was chosen as $\varphi (\lambda)=-{\lambda}log{\lambda}$. Each experiment was repeated for $10$ times and the results were averaged.

\begin{figure}[!t]
\centering
\begin{minipage}{4.1cm}
\centerline{
\includegraphics[width=1.8in]{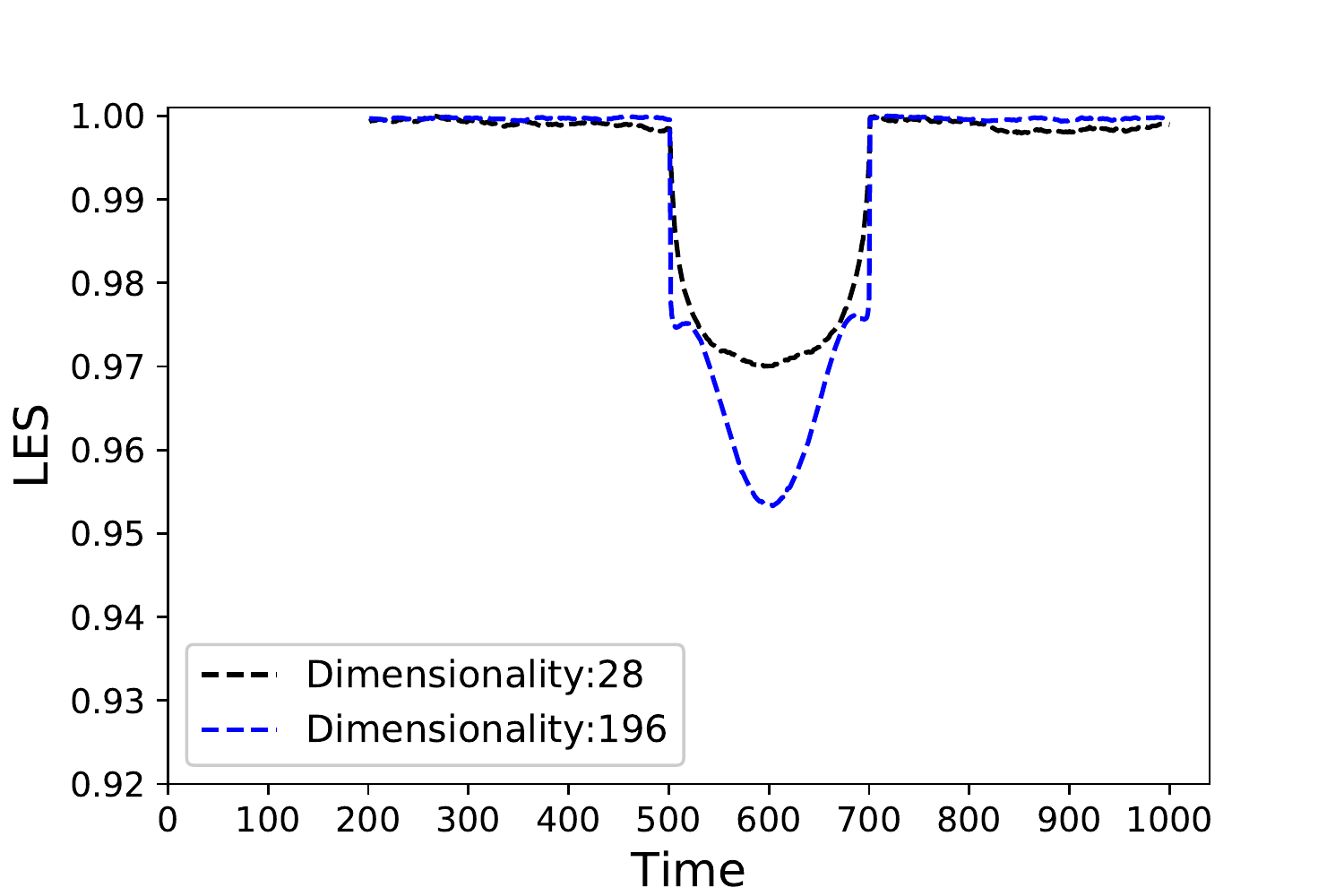}
}
\parbox{5cm}{\small \hspace{1.2cm}(a) $LES-t$ curve}
\end{minipage}
\hspace{0.2cm}
\begin{minipage}{4.1cm}
\centerline{
\includegraphics[width=1.8in]{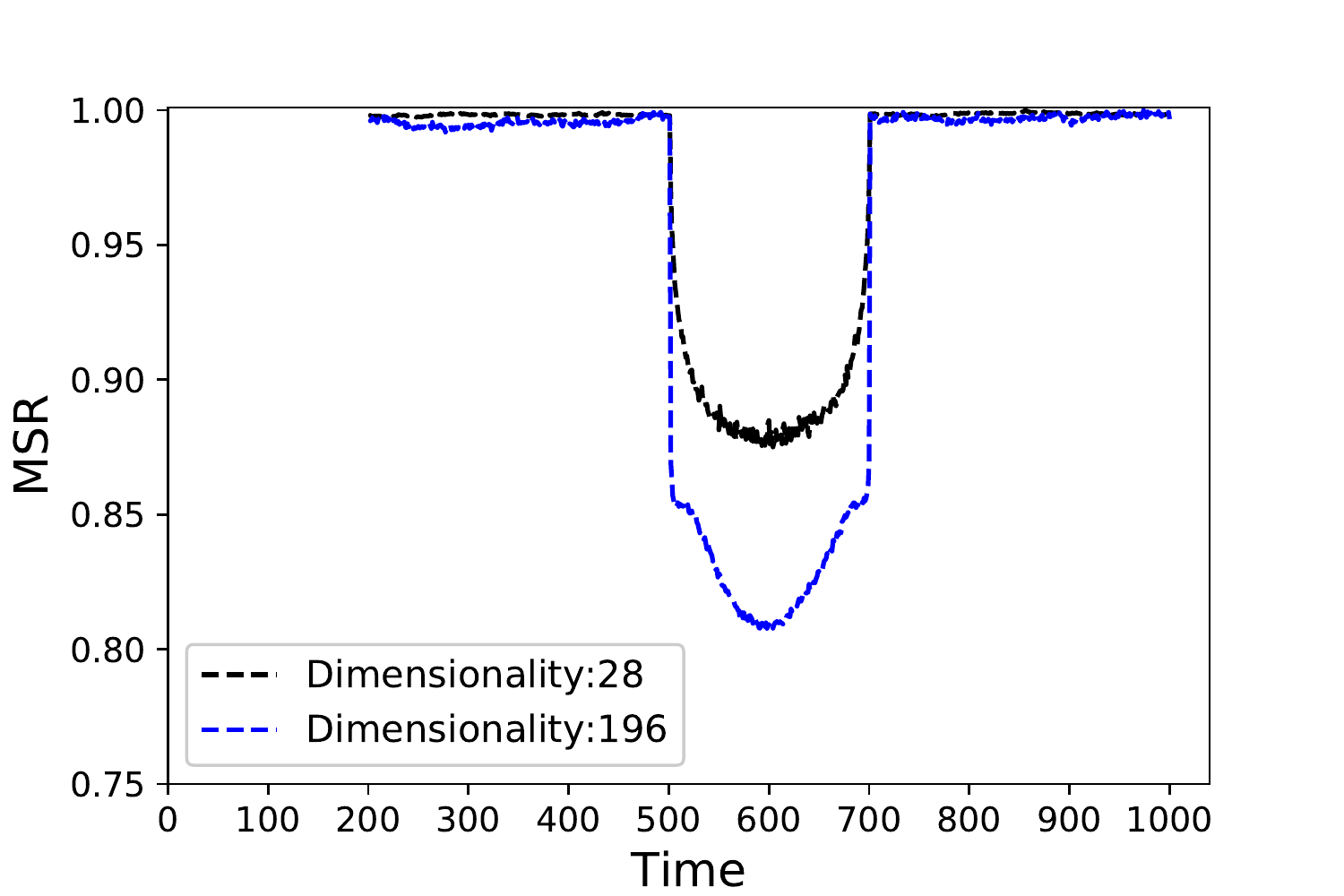}
}
\parbox{5cm}{\small \hspace{1.5cm}(b) $MSR-t$ curve}
\end{minipage}
\caption{The anomaly detection results of RMT based approach in Case A.}
\label{fig:case1_indicator}
\end{figure}
\begin{figure}[!t]
\centering
\begin{minipage}{4.1cm}
\centerline{
\includegraphics[width=1.8in]{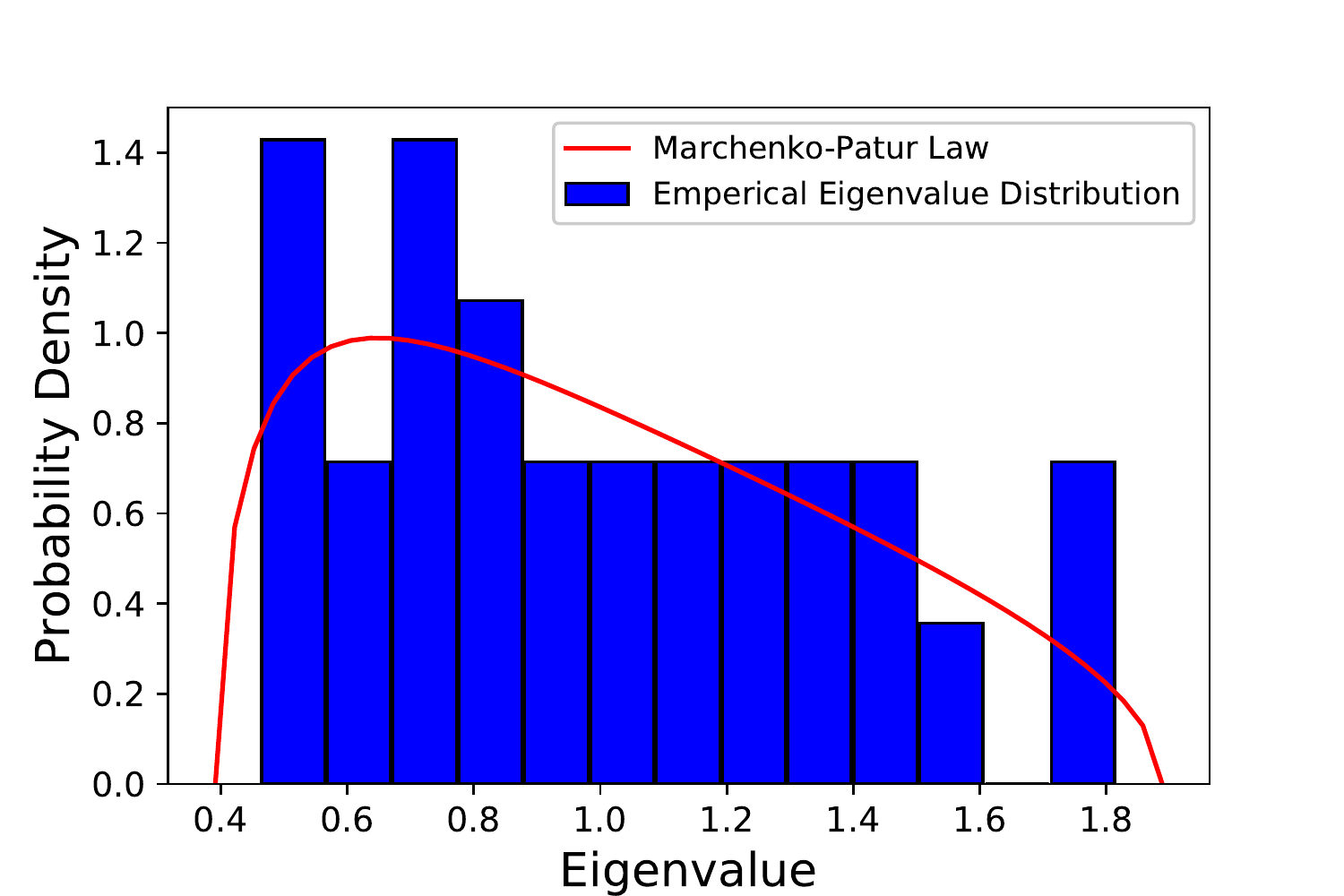}
}
\parbox{5cm}{\small \hspace{0.5cm}(a1)Dimensionality: $28$}
\end{minipage}
\hspace{0.2cm}
\begin{minipage}{4.1cm}
\centerline{
\includegraphics[width=1.8in]{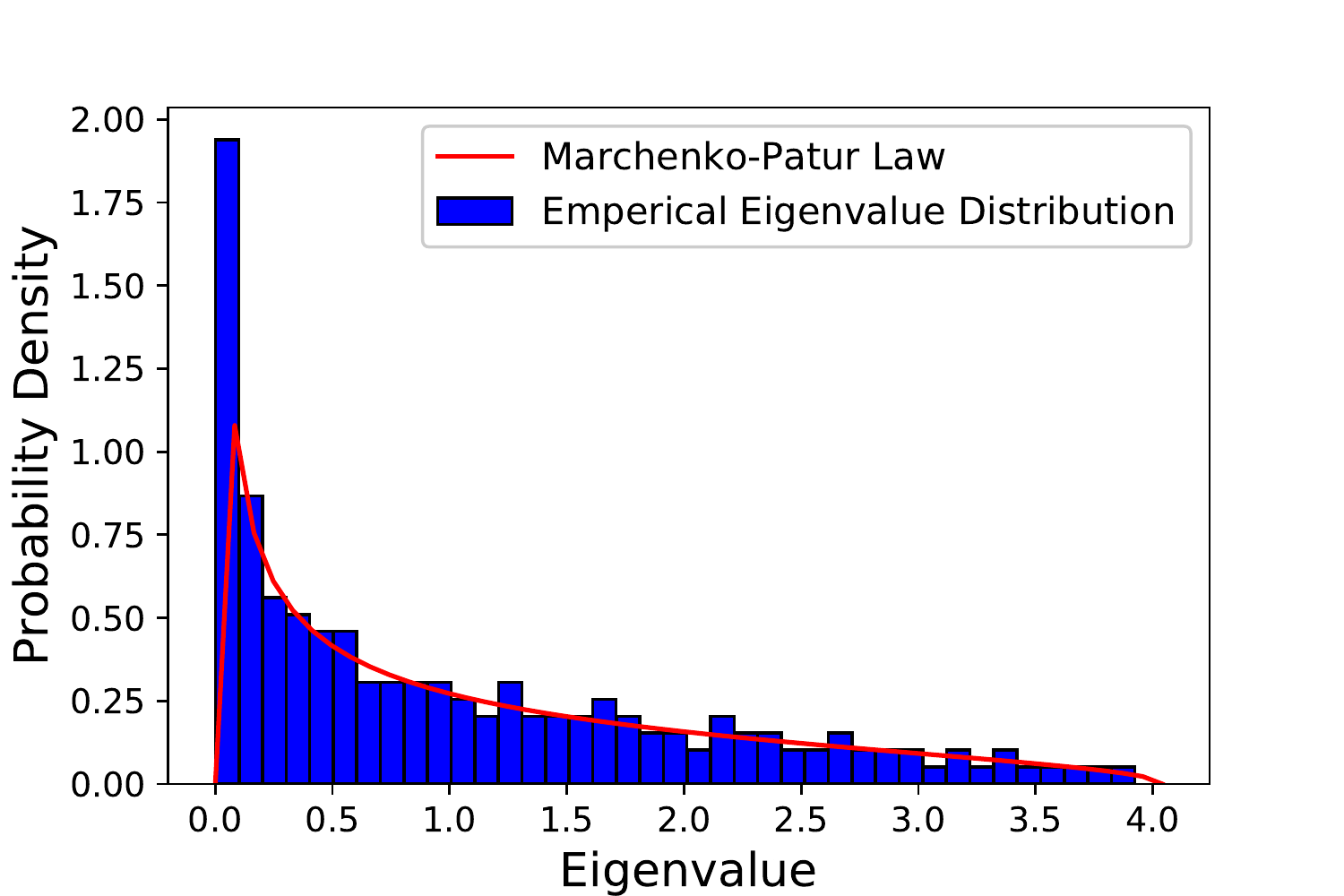}
}
\parbox{5cm}{\small \hspace{0.5cm}(a2)Dimensionality: $196$}
\end{minipage}
\hspace{0.2cm}
\begin{minipage}{4.1cm}
\centerline{
\includegraphics[width=1.8in]{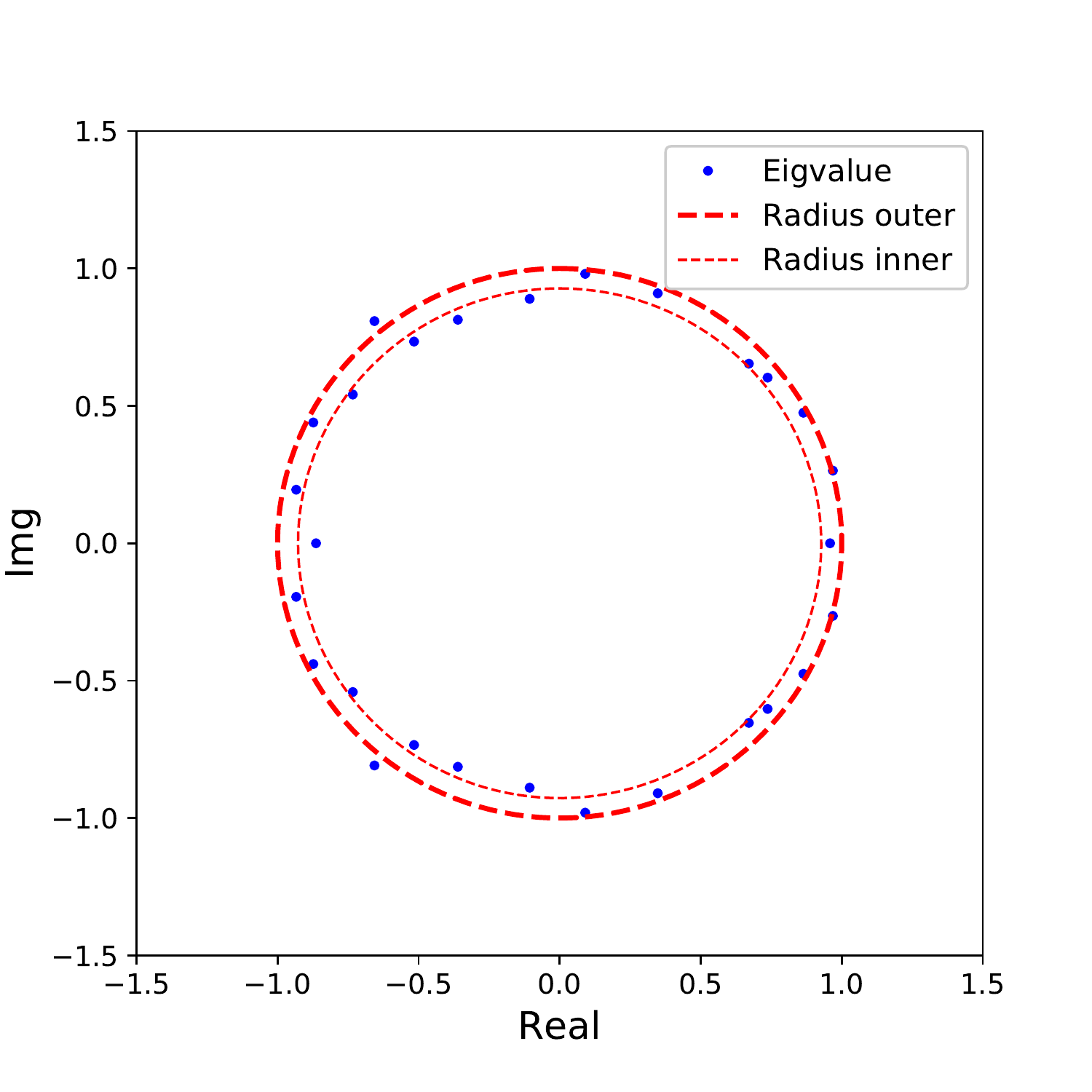}
}
\parbox{5cm}{\small \hspace{0.5cm}(b1)Dimensionality: $28$}
\end{minipage}
\hspace{0.2cm}
\begin{minipage}{4.1cm}
\centerline{
\includegraphics[width=1.8in]{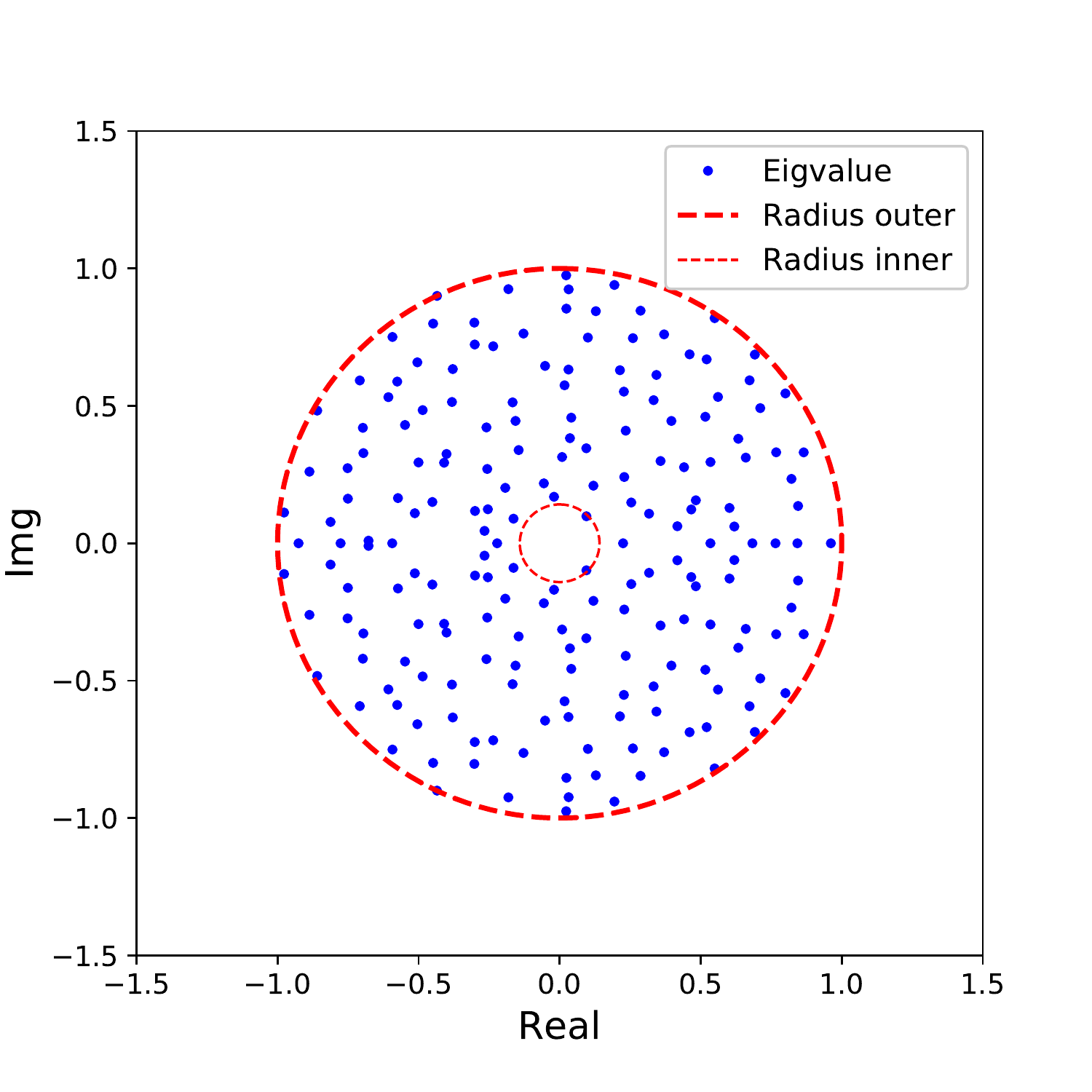}
}
\parbox{5cm}{\small \hspace{0.5cm}(b2)Dimensionality: $196$}
\end{minipage}
\caption{At $t_s=500$, the ESDs converge almost surely to the theoretical limits and they can converge better when the data dimensionality was increased from $28$ to $196$.}
\label{fig:case1_normal_law}
\end{figure}

The anomaly detection results are shown in Figure \ref{fig:case1_indicator}. The $LES-t$ curves in Figure \ref{fig:case1_indicator}(a) and $MSR-t$ curves in Figure \ref{fig:case1_indicator}(b) corresponding to different data dimensionality are normalized into $(0,1]$. It is noted that the curves begin at $t_s=200$, because the initial moving window includes $199$ times of historical sampling and the present sampling data. From the figures, it can be obtained:

\uppercase\expandafter{\romannumeral1}. From $t_s=200\sim 500$, $LES$ and $MSR$ corresponding to different data dimensionality remain almost constant, which indicates no anomalies occur and the system operates in steady state. Take $t_s=500$ for example, the ESDs converge almost surely to the theoretical M-P law and Ring law, which is shown in Figure \ref{fig:case1_normal_law}. It is observed that the ESDs converge to their theoretical limits better when the data dimensionality was increased from $28$ to $196$.

\begin{figure}[!t]
\centering
\begin{minipage}{4.1cm}
\centerline{
\includegraphics[width=1.8in]{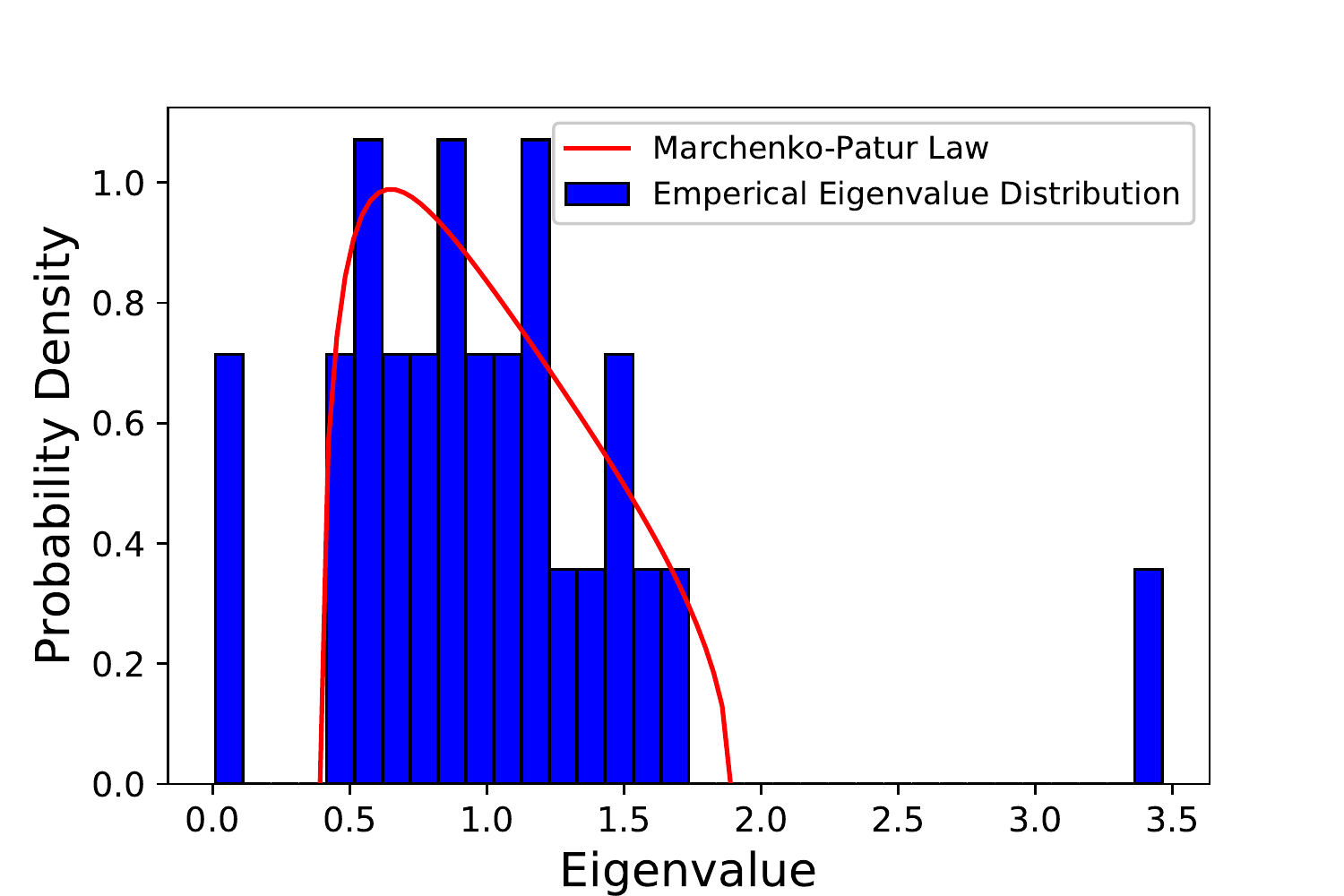}
}
\parbox{5cm}{\small \hspace{0.5cm}(a1)Dimensionality: $28$}
\end{minipage}
\hspace{0.2cm}
\begin{minipage}{4.1cm}
\centerline{
\includegraphics[width=1.8in]{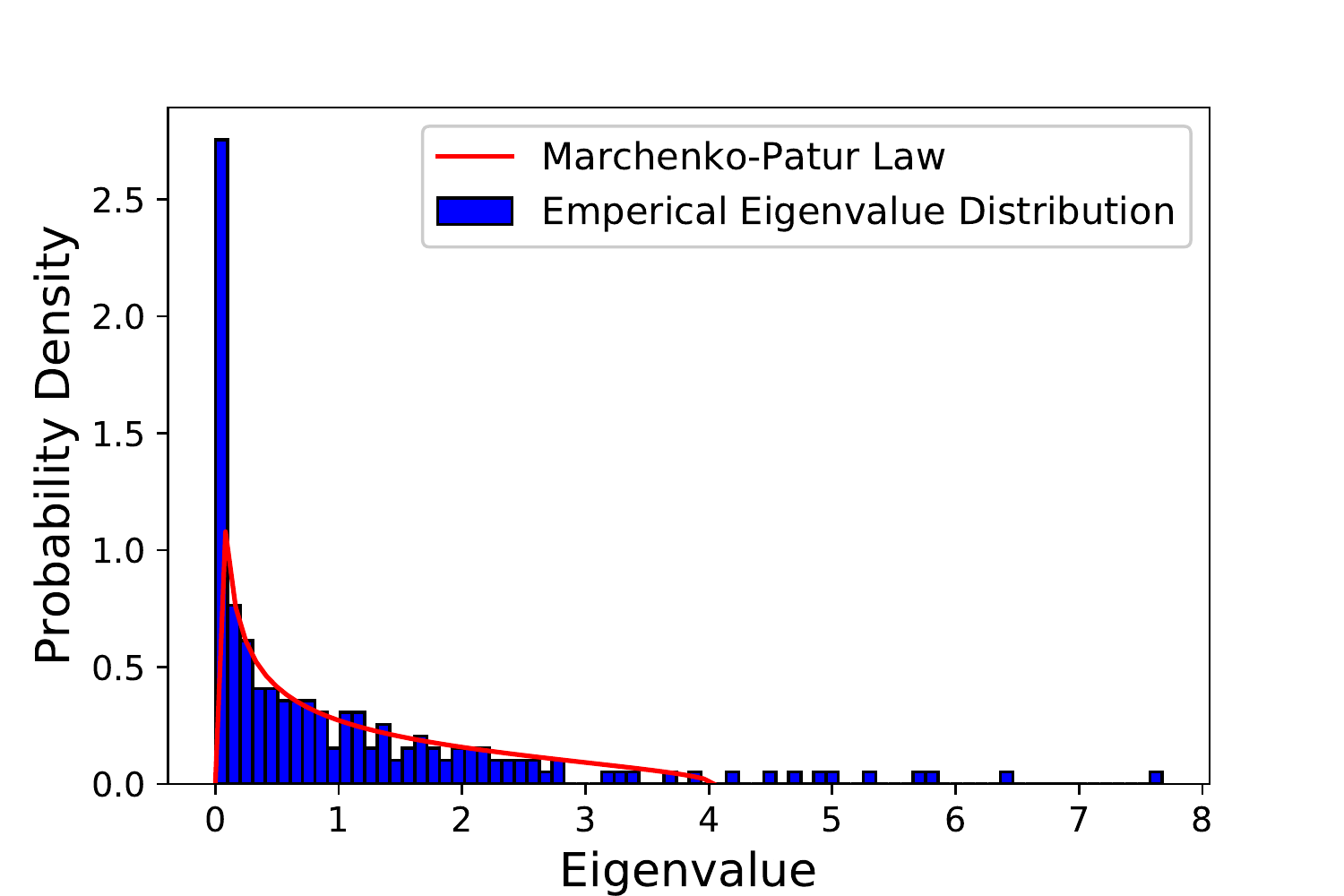}
}
\parbox{5cm}{\small \hspace{0.5cm}(a2)Dimensionality: $196$}
\end{minipage}
\hspace{0.2cm}
\begin{minipage}{4.1cm}
\centerline{
\includegraphics[width=1.8in]{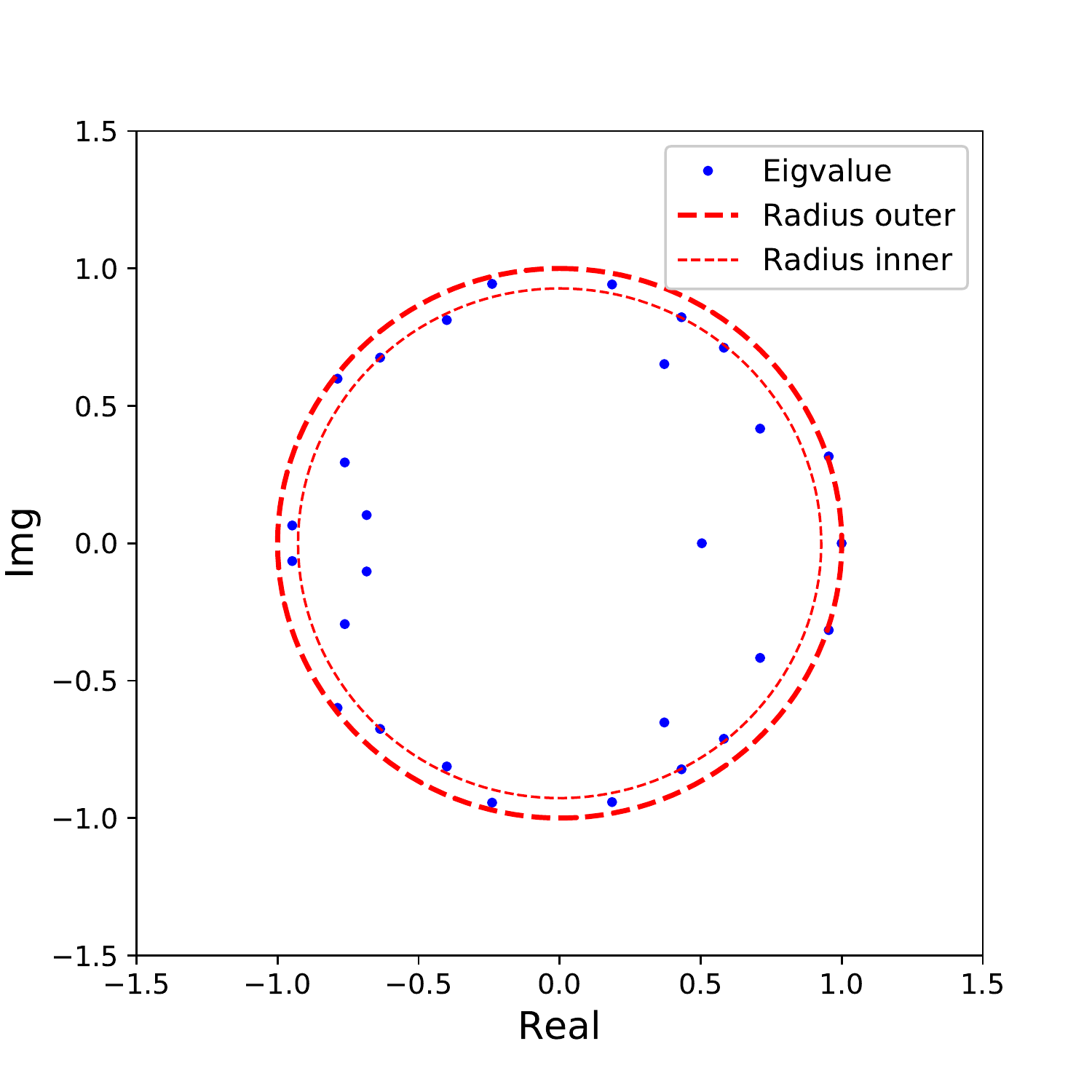}
}
\parbox{5cm}{\small \hspace{0.5cm}(b1)Dimensionality: $28$}
\end{minipage}
\hspace{0.2cm}
\begin{minipage}{4.1cm}
\centerline{
\includegraphics[width=1.8in]{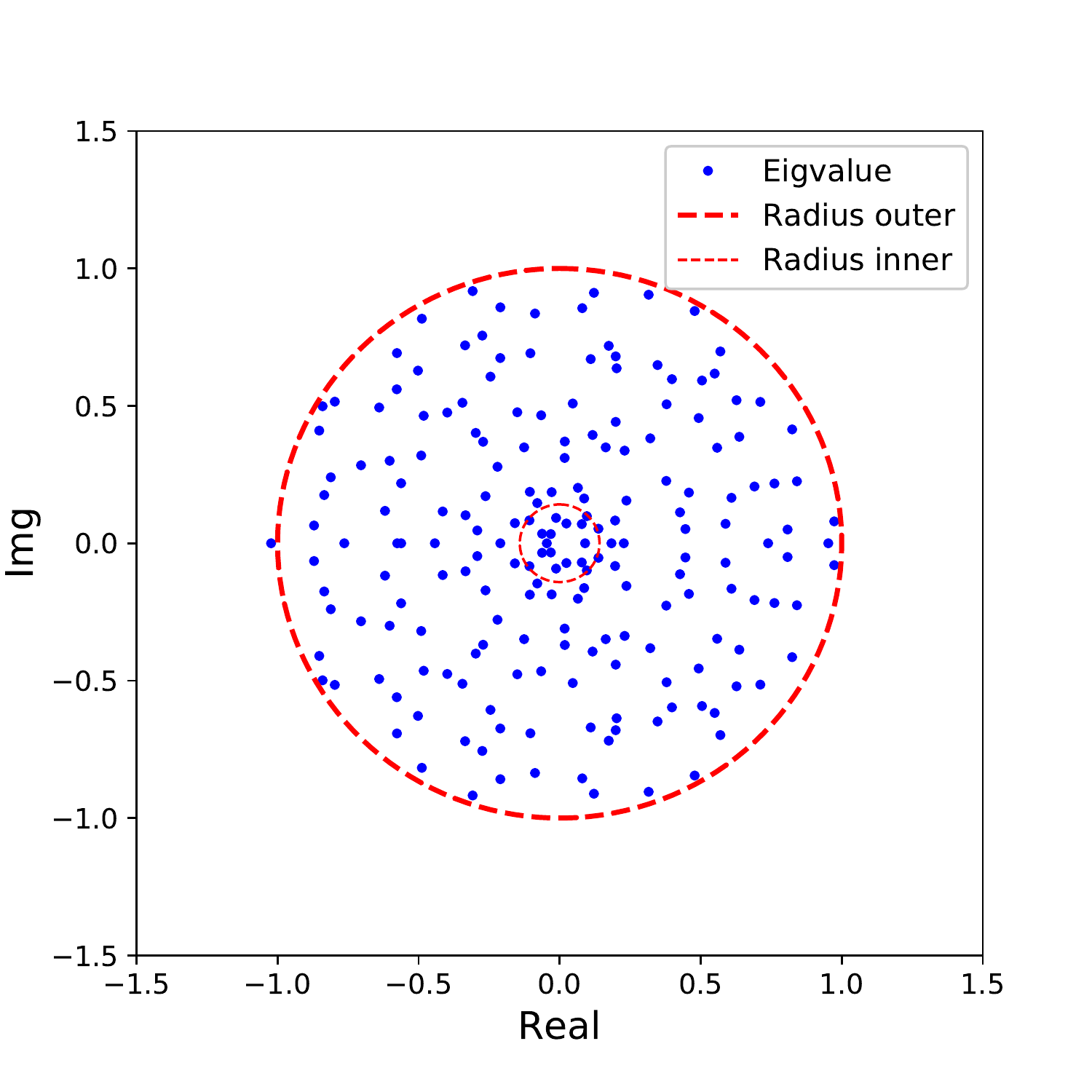}
}
\parbox{5cm}{\small \hspace{0.5cm}(b2)Dimensionality: $196$}
\end{minipage}
\caption{At $t_s=501$, the ESDs do not converge to the theoretical limits. More outliers occur and deviate further from the theoretical limits when the data dimensionality was increased from $28$ to $196$.}
\label{fig:case1_abnormal_law}
\end{figure}

\uppercase\expandafter{\romannumeral2}. At $t_s=501$, $LES$ and $MSR$ change rapidly, which indicates an anomaly is detected and the system operates in unsteady state. It is noted that the $LES-t$ and $MSR-t$ curves corresponding to high data dimensionality ($196$) have greater variance ratio, which demonstrates the proposed increasing data dimensionality algorithm can help improve the detection sensitivity of RMT based anomaly detection approach.
From $t_s=501\sim 700$, the $LES-t$ and $MSR-t$ curves are almost U-shaped, because the delay lag of the anomaly signal to $LES$ or $MSR$ is equal to the moving window's width. Take $t_s=501$ for example, the ESDs do not converge to the theoretical M-P law and Ring law, which is shown in Figure \ref{fig:case1_abnormal_law}. More outliers occur and deviate further from the theoretical limits when the data dimensionality was increased from $28$ to $196$.

\uppercase\expandafter{\romannumeral3}. From $t_s=701$, $LES$ and $MSR$ return to normal and remain constant afterwards, which indicates the anomaly signal does not contain in the moving window any more and the system operates in steady state.

2) ML Based Anomaly Detection Approach: The effectiveness of the proposed increasing data dimensionality for ML based anomaly detection approach has been explored and discussed in this case. The current ML based anomaly detection approach, such as SAE, has been well studied in \cite{sakurada2014anomaly,liu2018anomaly}. The synthetic data in Figure \ref{fig:case1_data_org} was used, and the dimensionality of the data was increased from $28$ to $196$ by using the proposed dimensionality increment algorithm. In the experiments, we analyzed both the original data and the dimensionality increment version through the SAE based anomaly detection approach. We trained the SAE prediction model by using a normal data sequence during $t_s=1\sim 200$, and computed the prediction errors for the remaining sequence during $t_s=201\sim 1000$, in which one time of sampling was used as a data sample. The parameters involved in the SAE approach are summarized as follows: \\
--the model depth: $3$; \\
--the number of neurons in each layer of encoder: \\
$28/196,48,24$; \\
--the number of neurons in each layer of decoder: \\
$24,48,28/196$; \\
--the initial learning rate: $0.0001$;  \\
--the activation function: $sigmoid$; \\
--the maximum iterations: $1000$;  \\
--the optimizer: $Adam$.  \\

\begin{figure}[!t]
\centerline{
\includegraphics[width=2.5in]{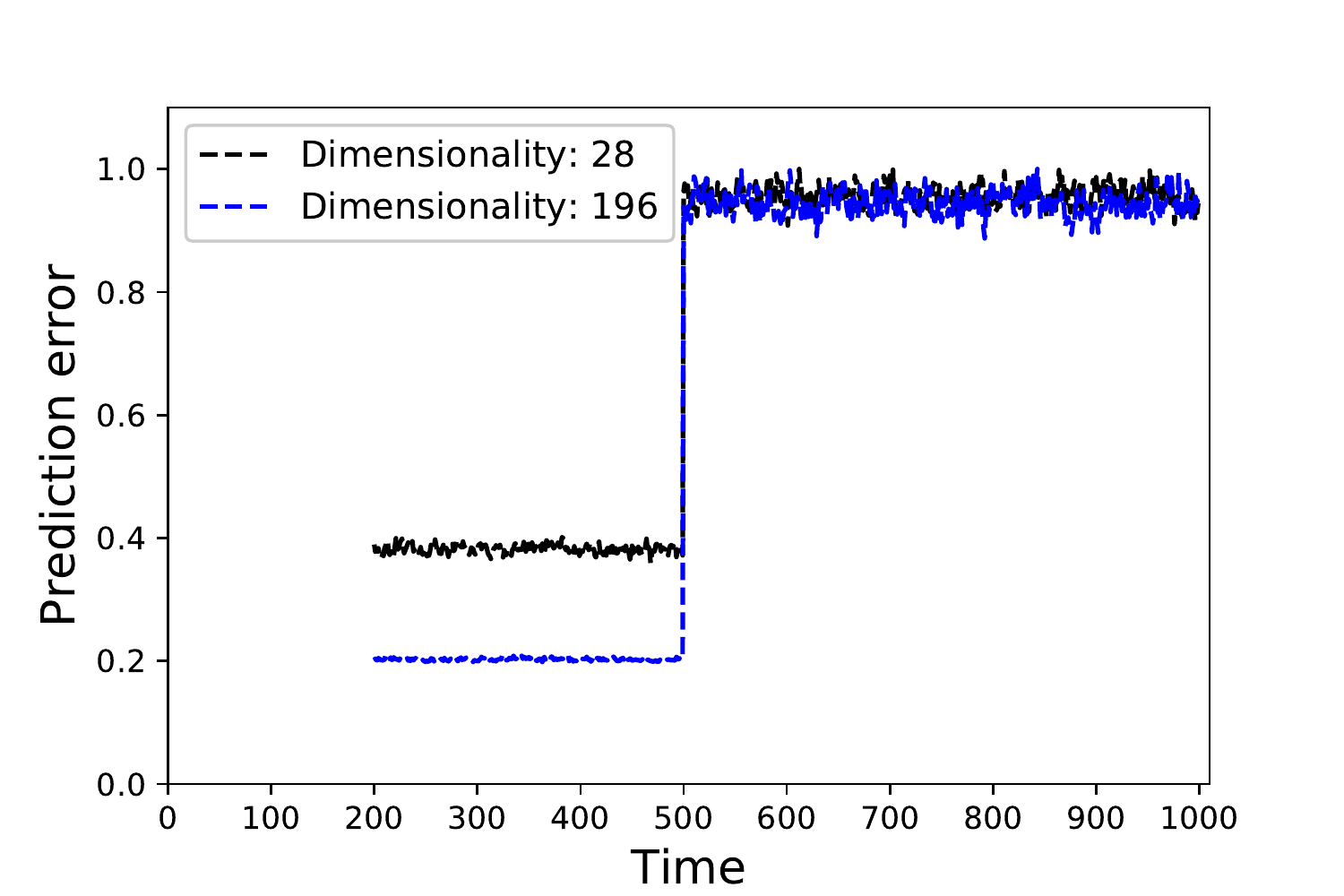}
}
\caption{The anomaly detection results of SAE based approach in Case A.}
\label{fig:case2_accuracy}
\end{figure}
Sensitivity Analysis: Figure \ref{fig:case2_accuracy} shows the anomaly detection results of SAE based approach for different data dimensionality. The results were normalized into $[0,1]$. It can be observed the prediction error curves change dramatically at $t_s=501$, which indicates the anomaly is detected effectively. What's more, the blue dashed line corresponding to high data dimensionality has a higher variance ratio at $t_s=501$, which indicates it becomes easier for detecting the anomaly when the data's dimensionality was increased from $28$ to $196$ through our proposed algorithm.

\begin{figure}[!t]
\centerline{
\includegraphics[width=2.5in]{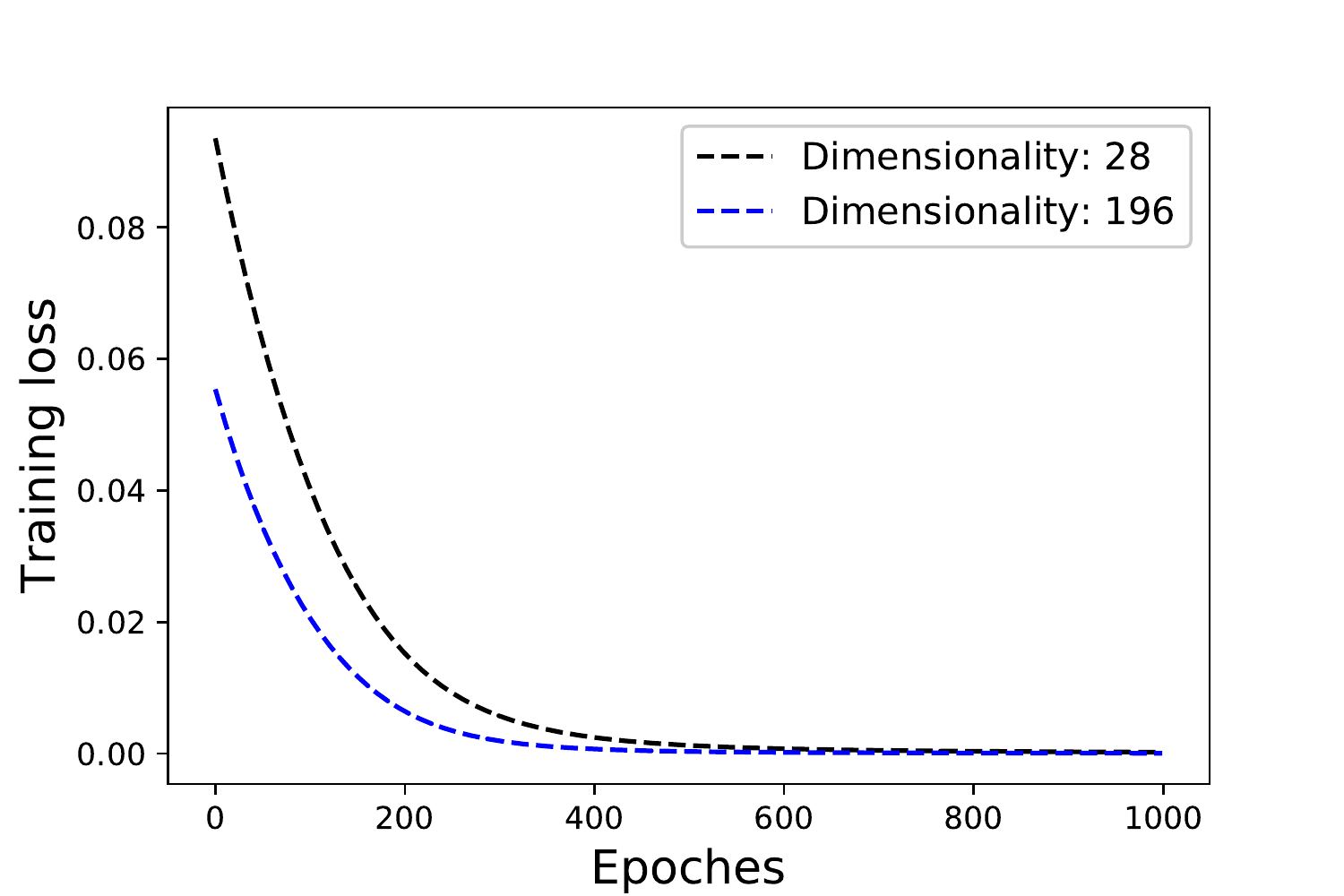}
}
\caption{The convergence rate of SAE based approach in Case A.}
\label{fig:case2_convergence}
\end{figure}
Convergence Analysis: Figure \ref{fig:case2_convergence} illustrates the convergence rate in training SAE for different data dimensionality. It can be observed that the blue dashed line corresponding to high data dimensionality converges faster, which validates the proposed algorithm can help improve the convergence rate of model training in SAE based anomaly detection approach.
\subsection{Case Studies on the Detection of Overload}
\label{subsection: case_B}
\begin{figure}[!t]
\centerline{
\includegraphics[width=2.5in]{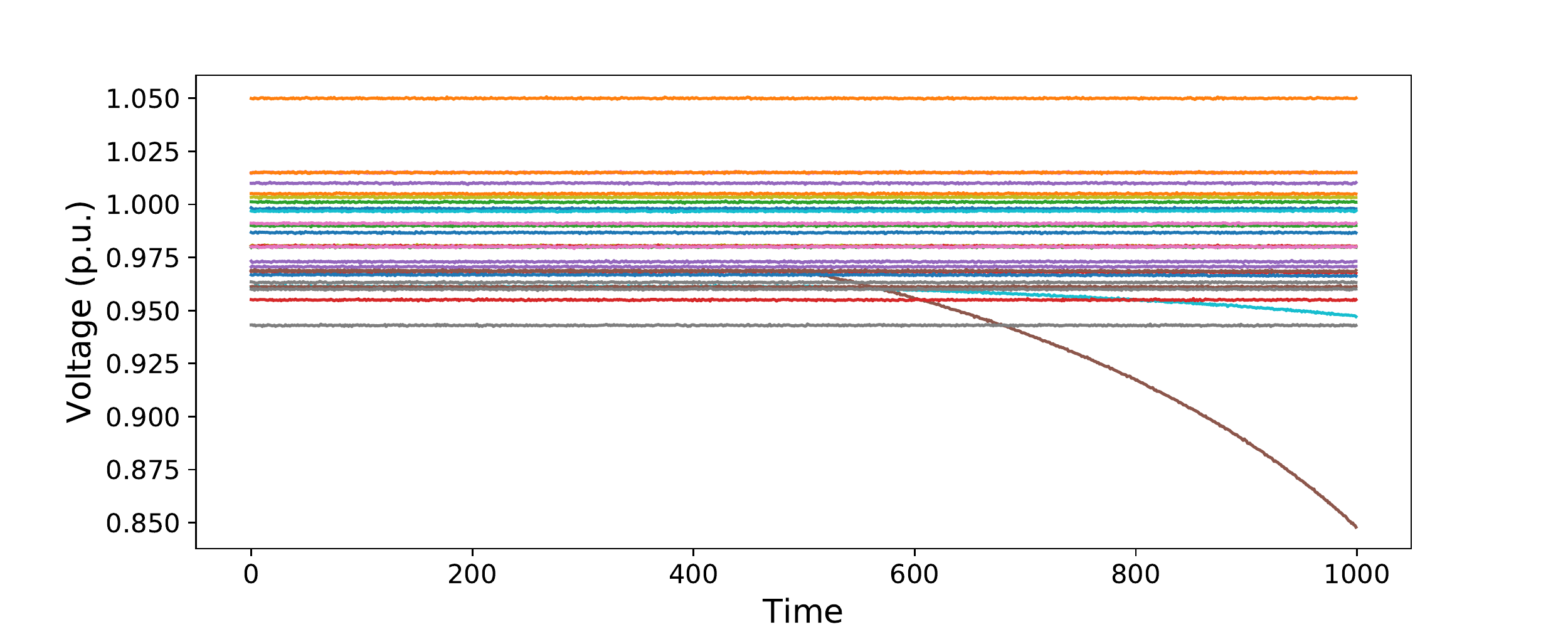}
}
\caption{The synthetic data collected from $28$ PMUs installed in the IEEE 118-bus test system in Case B. The overload anomaly was set from $t_s=501$.}
\label{fig:case3_data_org}
\end{figure}

In this section, the overload anomaly was set by gradually increasing the load at bus $20$ and others stayed unchanged, as shown in Table \ref{Tab: Case3}. The synthetic data collected from $28$ PMUs installed in the system was shown in Figure \ref{fig:case3_data_org}. It contained $28$ voltage measurement variables with sampling $1000$ times.
\begin{table}[!t]
\caption{The Anomaly of Overload at Bus $20$ in Case B.}
\label{Tab: Case3}
\centering
\begin{tabular}{clc}   
\toprule[1.0pt]
\textbf {Bus} & \textbf{Sampling Time}& \textbf{Active Power(MW)}\\
\hline
\multirow{2}*{20} & $t_s=1\sim 500$ & $20$ \\
~&$t_s=501\sim 1000$ & $20\rightarrow 320$ \\
\hline
Others & $t_s=1\sim 1000$ & Unchanged \\
\hline
\end{tabular}
\end{table}

1) RMT Based Anomaly Detection Approach: Similar in Case A.1, in this case, we test the effectiveness of the proposed increasing data dimensionality for RMT based anomaly detection approach on the generated synthetic data in Figure \ref{fig:case3_data_org}. The parameters involved in the experiments were set the same as in Case A.1. Each experiment was repeated for $10$ times and the results were averaged. The anomaly detection results corresponding to different data dimensionality are shown in Figure \ref{fig:case3_indicator}, in which the $LES-t$ curves and $MSR-t$ curves are normalized into $(0,1]$. From the figure, it can be obtained:
\begin{figure}[!t]
\centering
\begin{minipage}{4.1cm}
\centerline{
\includegraphics[width=1.8in]{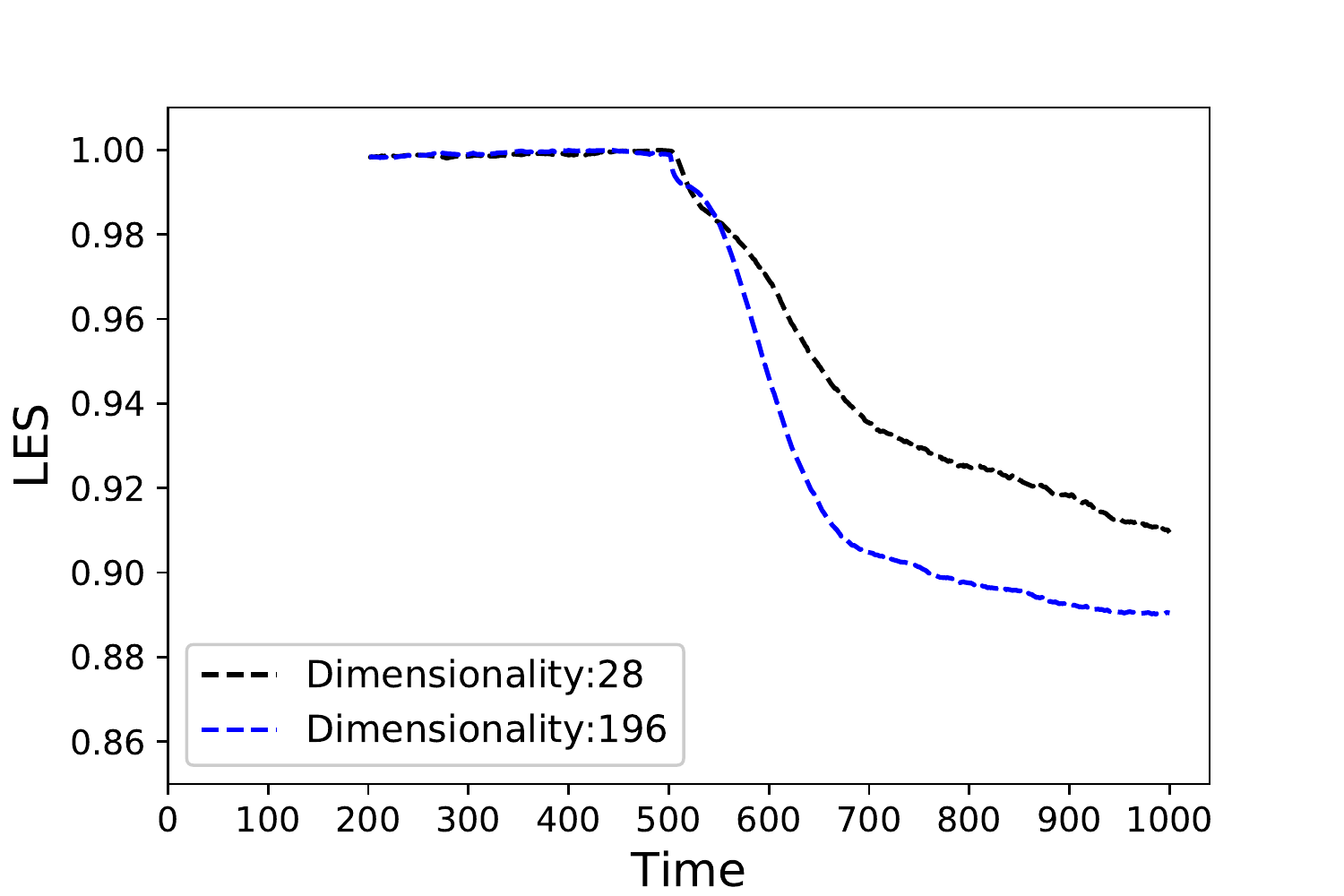}
}
\parbox{5cm}{\small \hspace{1.2cm}(a) $LES-t$ curve}
\end{minipage}
\hspace{0.2cm}
\begin{minipage}{4.1cm}
\centerline{
\includegraphics[width=1.8in]{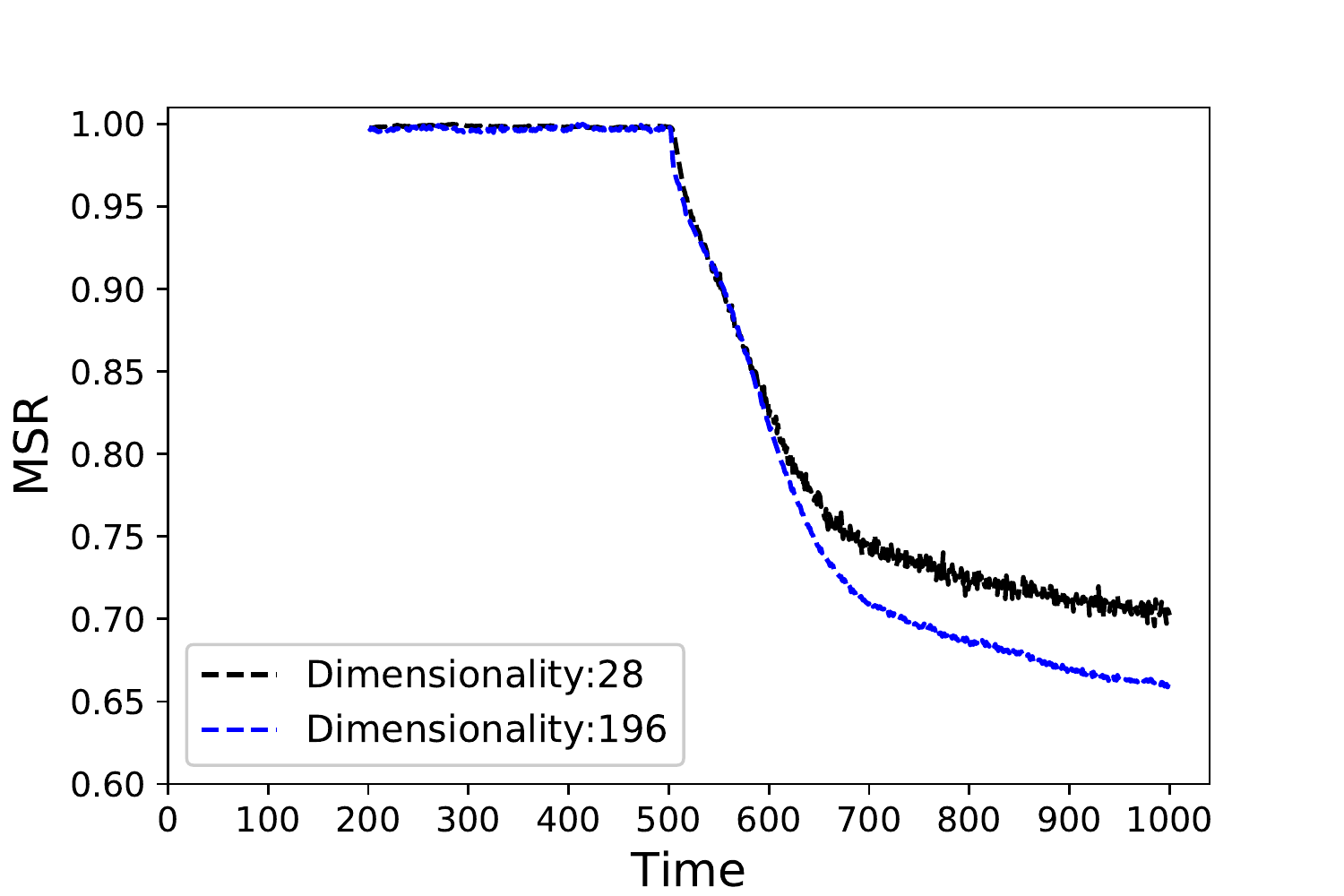}
}
\parbox{5cm}{\small \hspace{1.5cm}(b) $MSR-t$ curve}
\end{minipage}
\caption{The anomaly detection results of RMT based approach in Case B.}
\label{fig:case3_indicator}
\end{figure}

\begin{figure}[!t]
\centering
\begin{minipage}{4.1cm}
\centerline{
\includegraphics[width=1.8in]{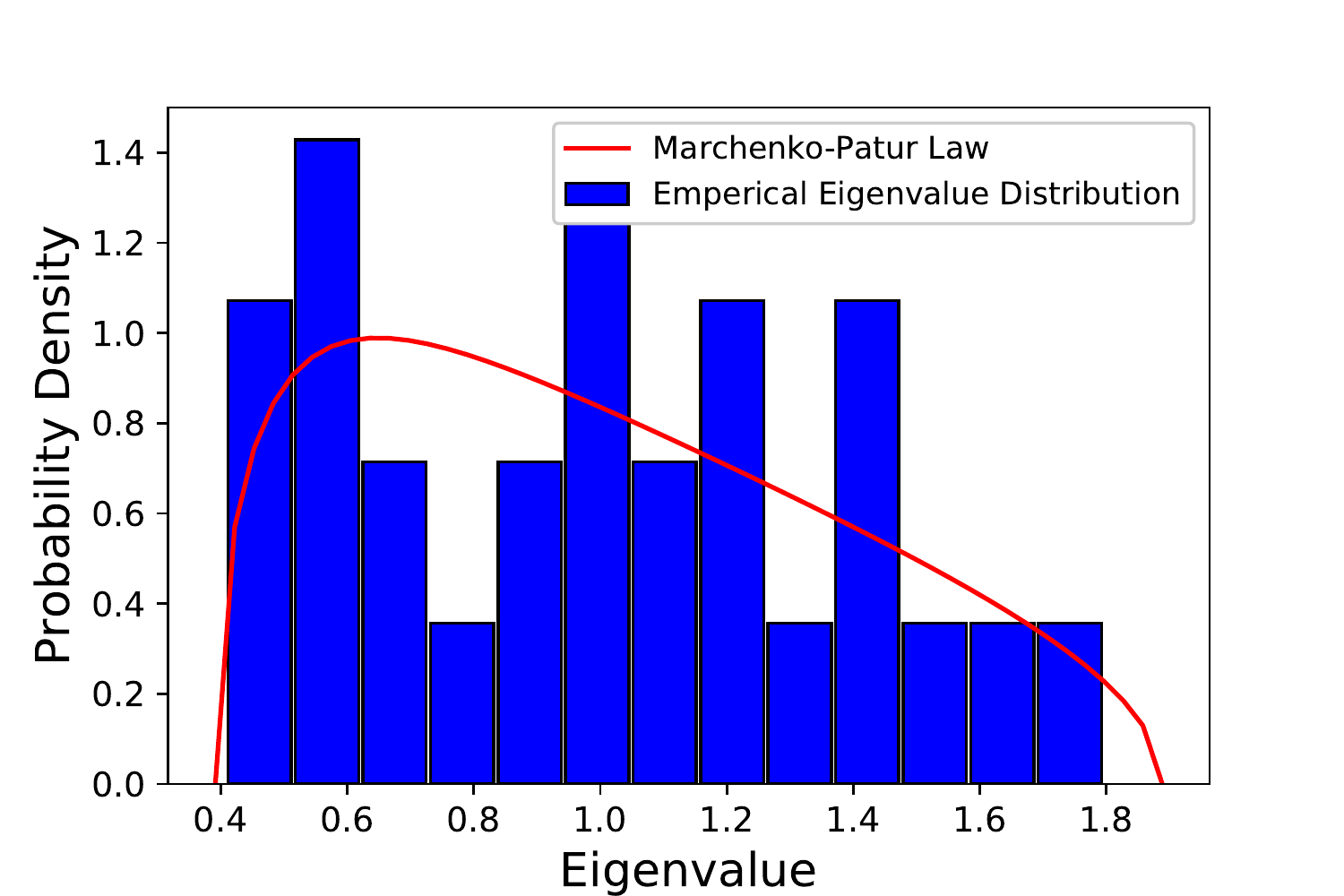}
}
\parbox{5cm}{\small \hspace{0.5cm}(a1)Dimensionality: $28$}
\end{minipage}
\hspace{0.2cm}
\begin{minipage}{4.1cm}
\centerline{
\includegraphics[width=1.8in]{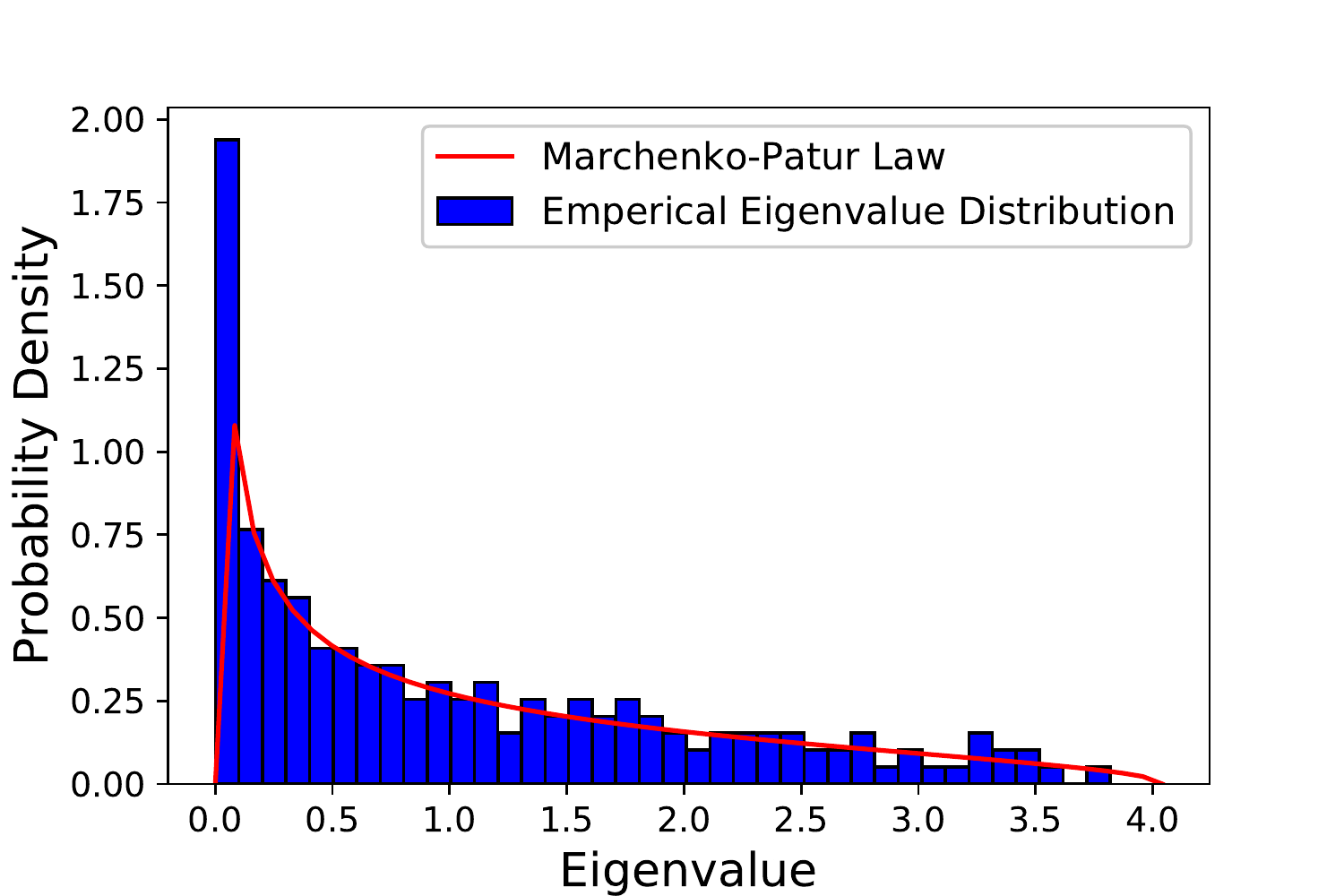}
}
\parbox{5cm}{\small \hspace{0.5cm}(a2)Dimensionality: $196$}
\end{minipage}
\hspace{0.2cm}
\begin{minipage}{4.1cm}
\centerline{
\includegraphics[width=1.8in]{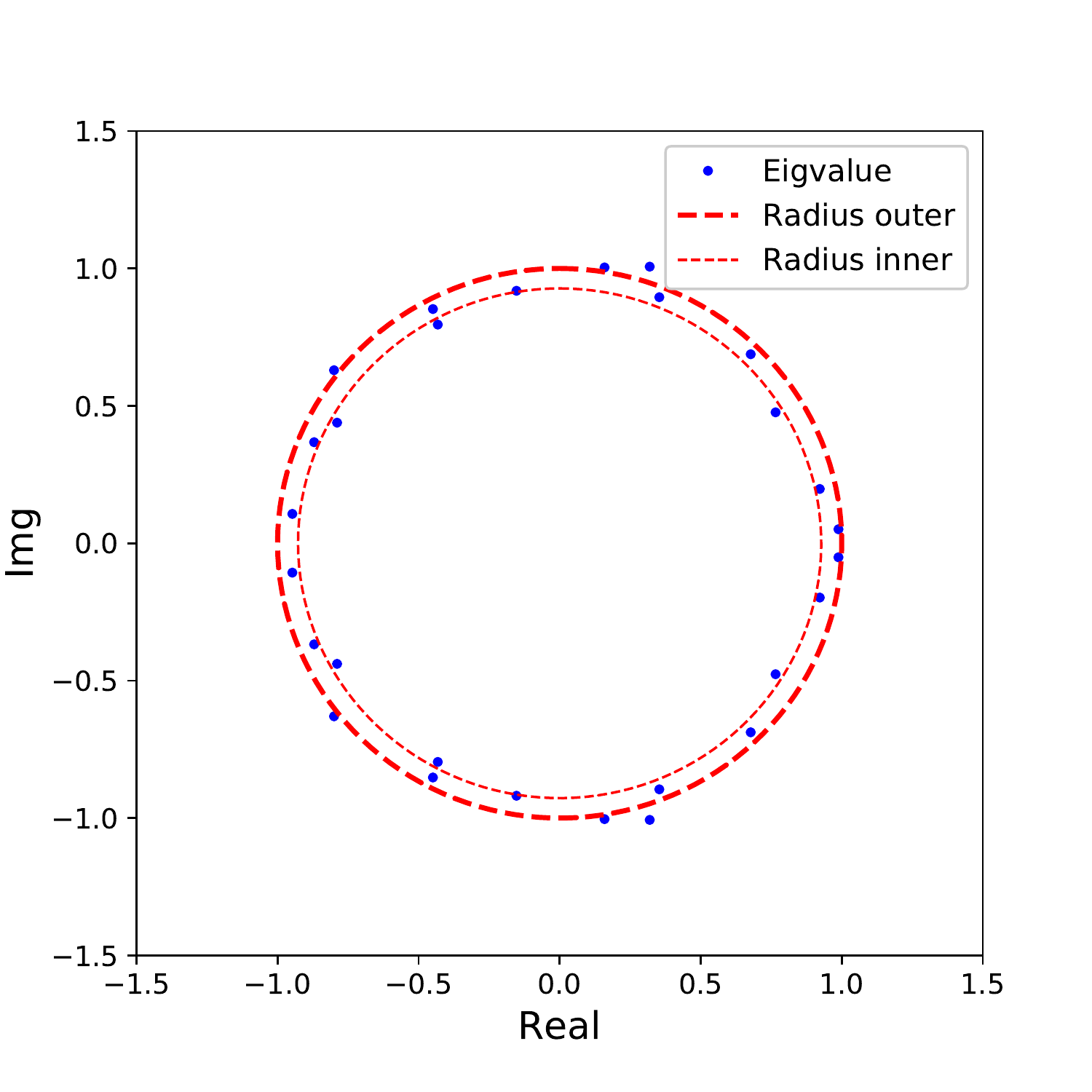}
}
\parbox{5cm}{\small \hspace{0.5cm}(b1)Dimensionality: $28$}
\end{minipage}
\hspace{0.2cm}
\begin{minipage}{4.1cm}
\centerline{
\includegraphics[width=1.8in]{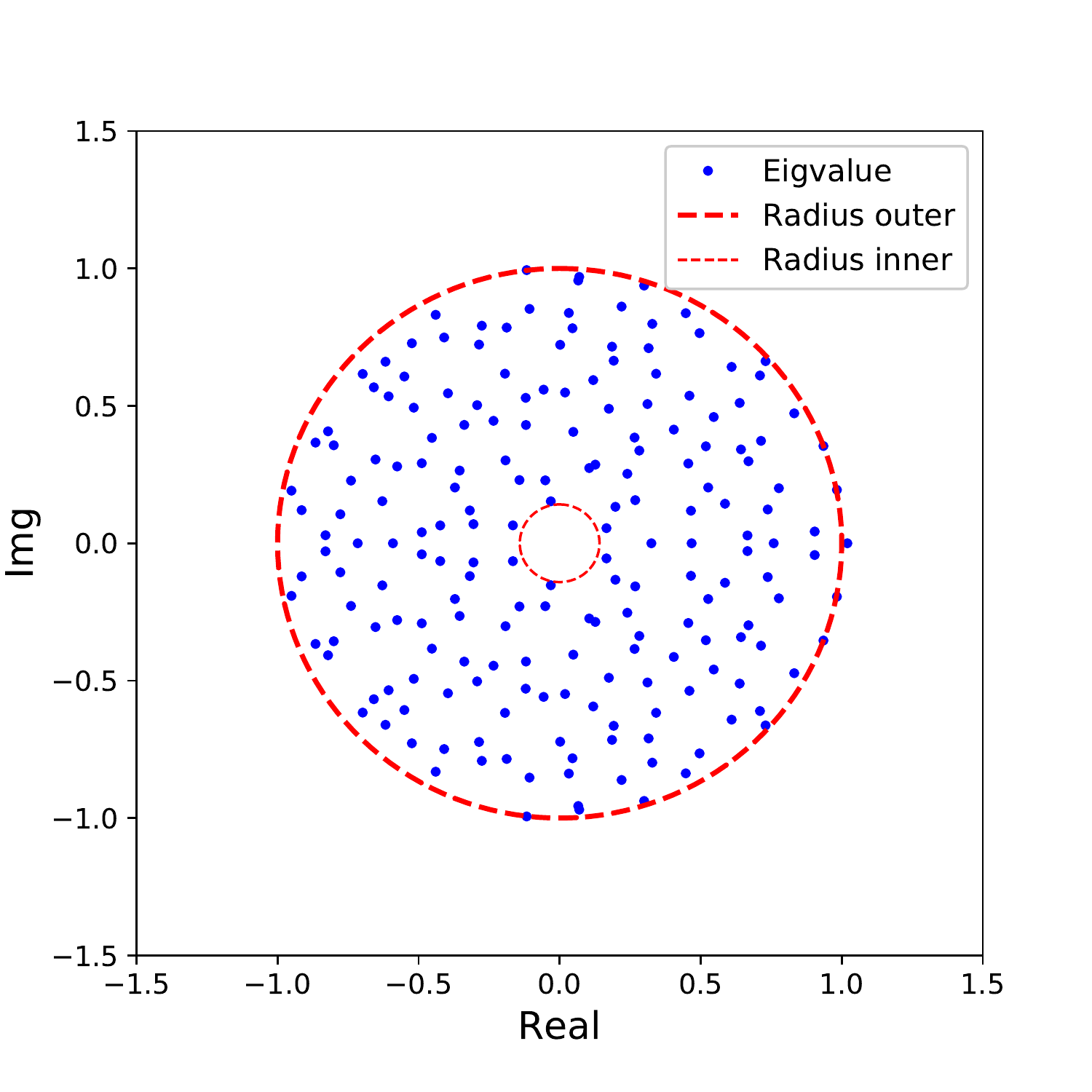}
}
\parbox{5cm}{\small \hspace{0.5cm}(b2)Dimensionality: $196$}
\end{minipage}
\caption{At $t_s=500$, the ESDs converge almost surely to the theoretical limits and they can converge better when the data dimensionality was increased from $28$ to $196$.}
\label{fig:case3_normal_law}
\end{figure}

\uppercase\expandafter{\romannumeral1}. During $t_s=200\sim 500$, the $LES$ and $MSR$ values corresponding to different data dimensionality remain almost constant, which indicates the system operates in steady state. For example, at $t_s=500$, the ESDs converge almost surely to the theoretical M-P law and Ring law, as shown in Figure \ref{fig:case3_normal_law}. It can be also observed that the ESDs converge to their theoretical limits better when the data dimensionality was increased from $28$ to $196$.

\begin{figure}[!t]
\centering
\begin{minipage}{4.1cm}
\centerline{
\includegraphics[width=1.8in]{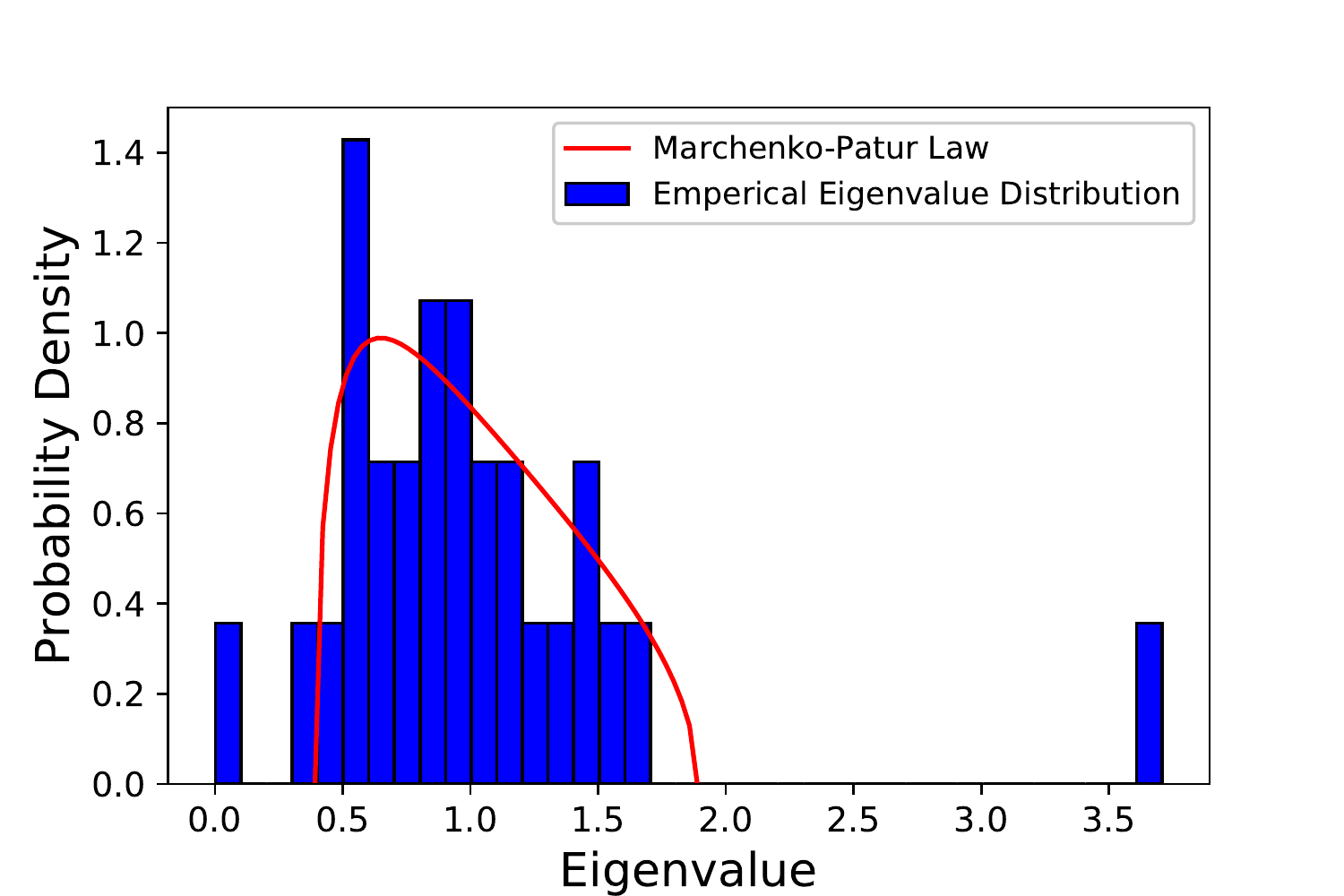}
}
\parbox{5cm}{\small \hspace{0.5cm}(a1)Dimensionality: $28$}
\end{minipage}
\hspace{0.2cm}
\begin{minipage}{4.1cm}
\centerline{
\includegraphics[width=1.8in]{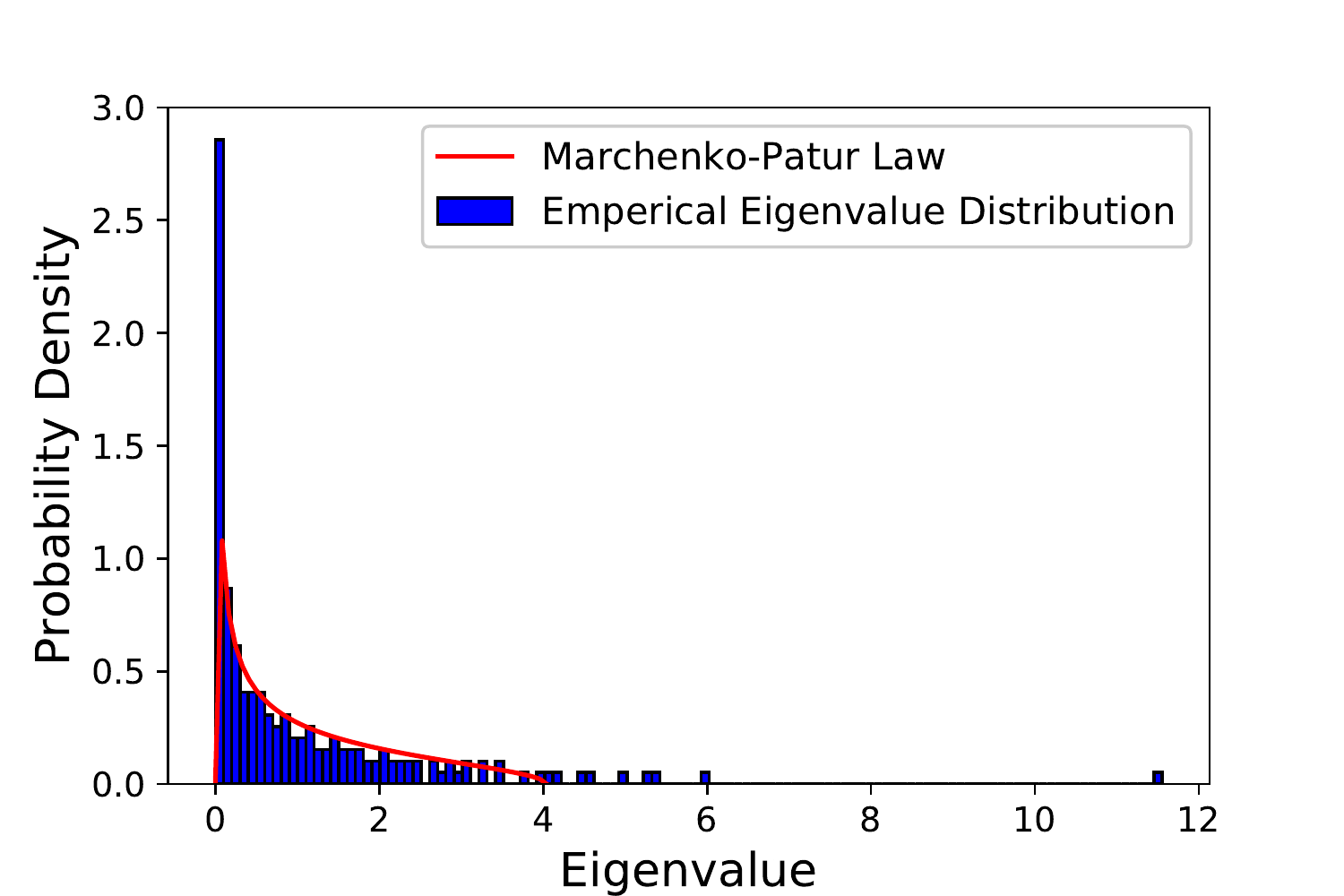}
}
\parbox{5cm}{\small \hspace{0.5cm}(a2)Dimensionality: $196$}
\end{minipage}
\hspace{0.2cm}
\begin{minipage}{4.1cm}
\centerline{
\includegraphics[width=1.8in]{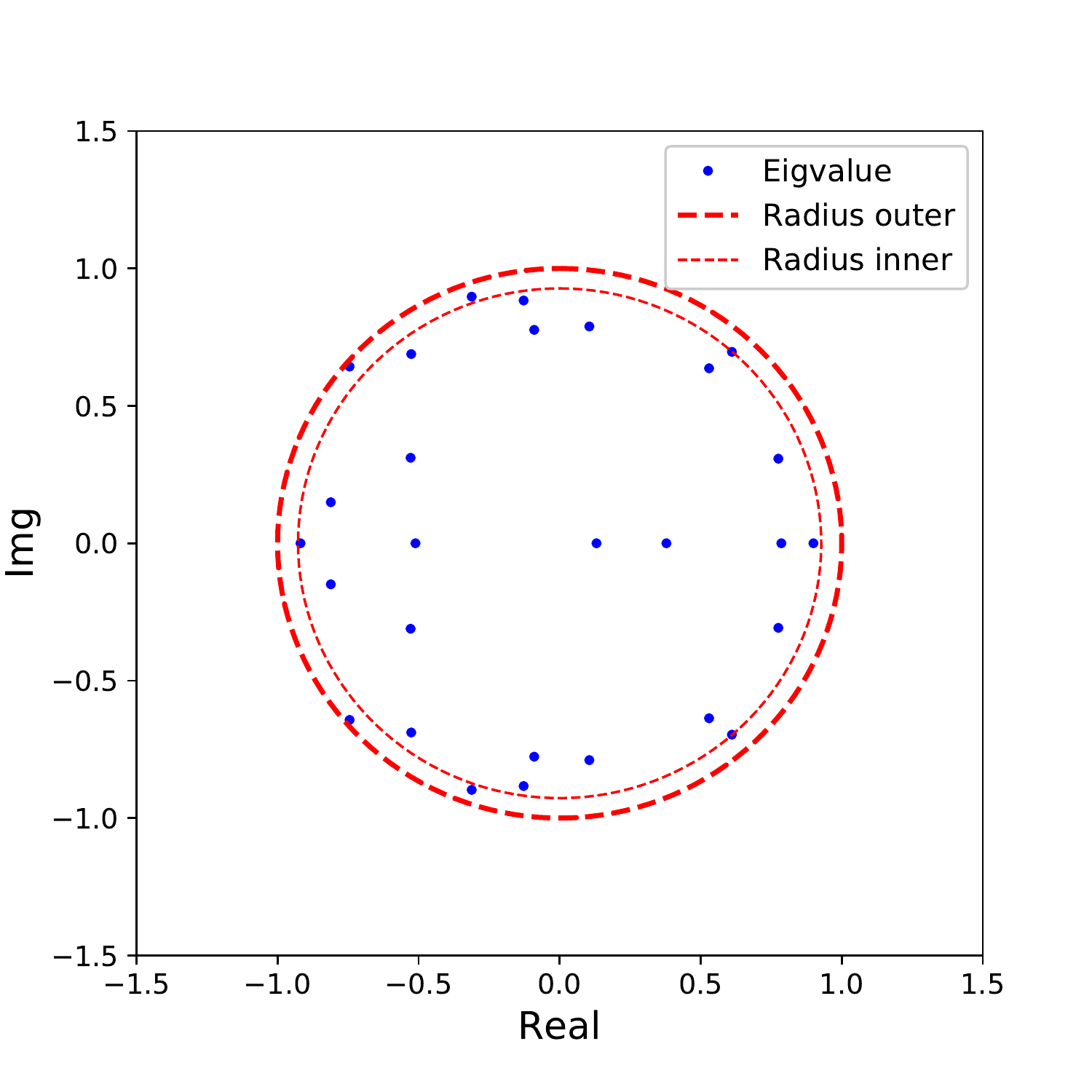}
}
\parbox{5cm}{\small \hspace{0.5cm}(b1)Dimensionality: $28$}
\end{minipage}
\hspace{0.2cm}
\begin{minipage}{4.1cm}
\centerline{
\includegraphics[width=1.8in]{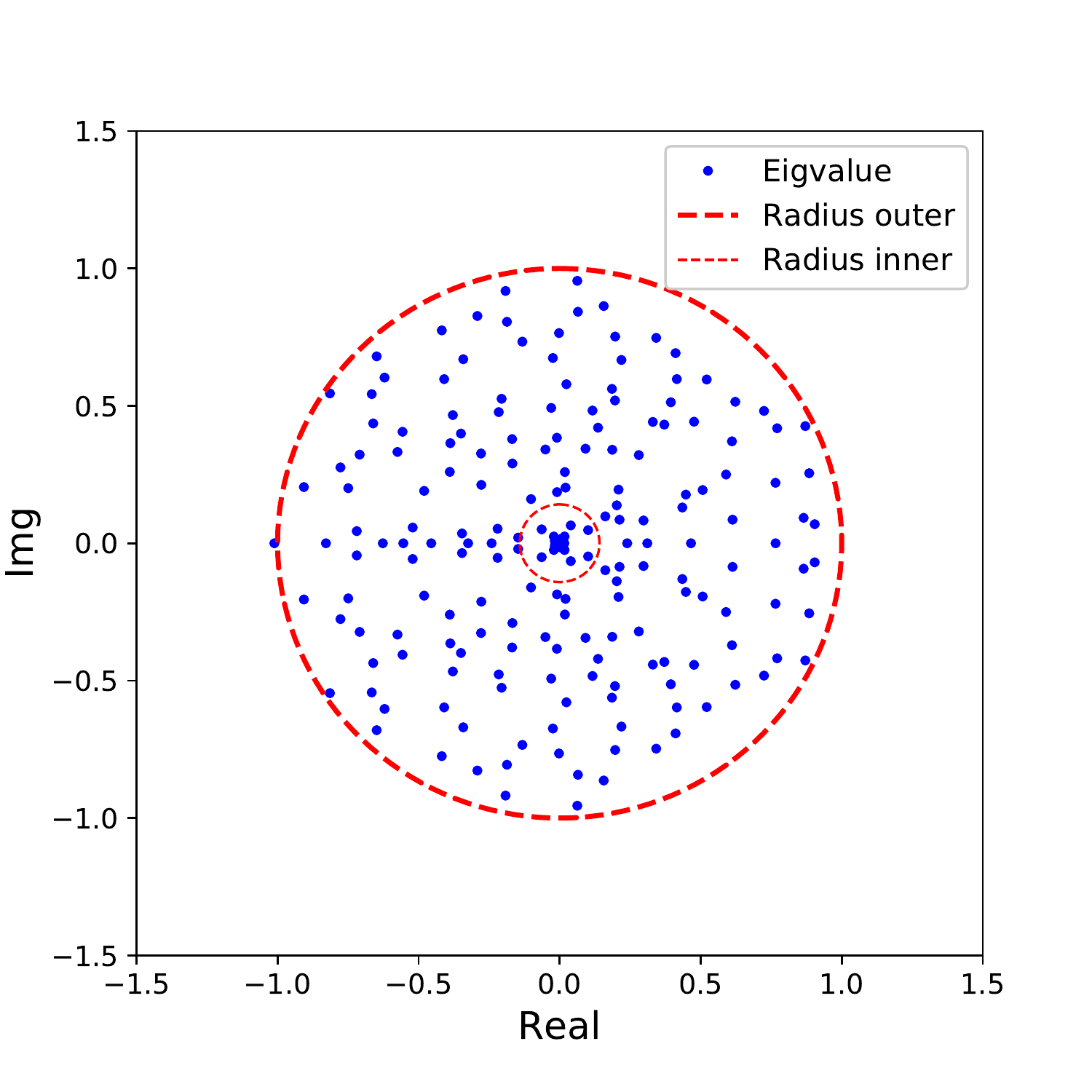}
}
\parbox{5cm}{\small \hspace{0.5cm}(b2)Dimensionality: $196$}
\end{minipage}
\caption{At $t_s=600$, the ESDs do not converge to the theoretical limits. More outliers occur and deviate further from the theoretical limits when the data dimensionality was increased from $28$ to $196$.}
\label{fig:case3_abnormal_law}
\end{figure}

\uppercase\expandafter{\romannumeral2}. From $t_s=501$, the $LES$ and $MSR$ values corresponding to different data dimensionality begin to decrease gradually, which indicates an anomaly occurs and the system begins to operate in unsteady state. It is noted that the $LES-t$ and $MSR-t$ curves corresponding to high data dimensionality have greater variance ratio, which validates the proposed increasing data dimensionality algorithm can help improve the sensitivity of RMT based approach for anomaly detection. Take $t_s=600$ for example, the ESDs do not converge to the theoretical M-P law and Ring law under the unsteady state, as shown in Figure \ref{fig:case3_abnormal_law}. More outliers occur and deviate further from their theoretical limits when the data dimensionality was increased from $28$ to $196$.

2) ML Based Anomaly Detection Approach: Similar in Case A.2, in this case, we test the effectiveness of the proposed increasing data dimensionality for SAE based anomaly detection approach on the synthetic data in Figure \ref{fig:case3_data_org}. The experiments and parameters were set the same as in Case A.2.

\begin{figure}[!t]
\centerline{
\includegraphics[width=2.5in]{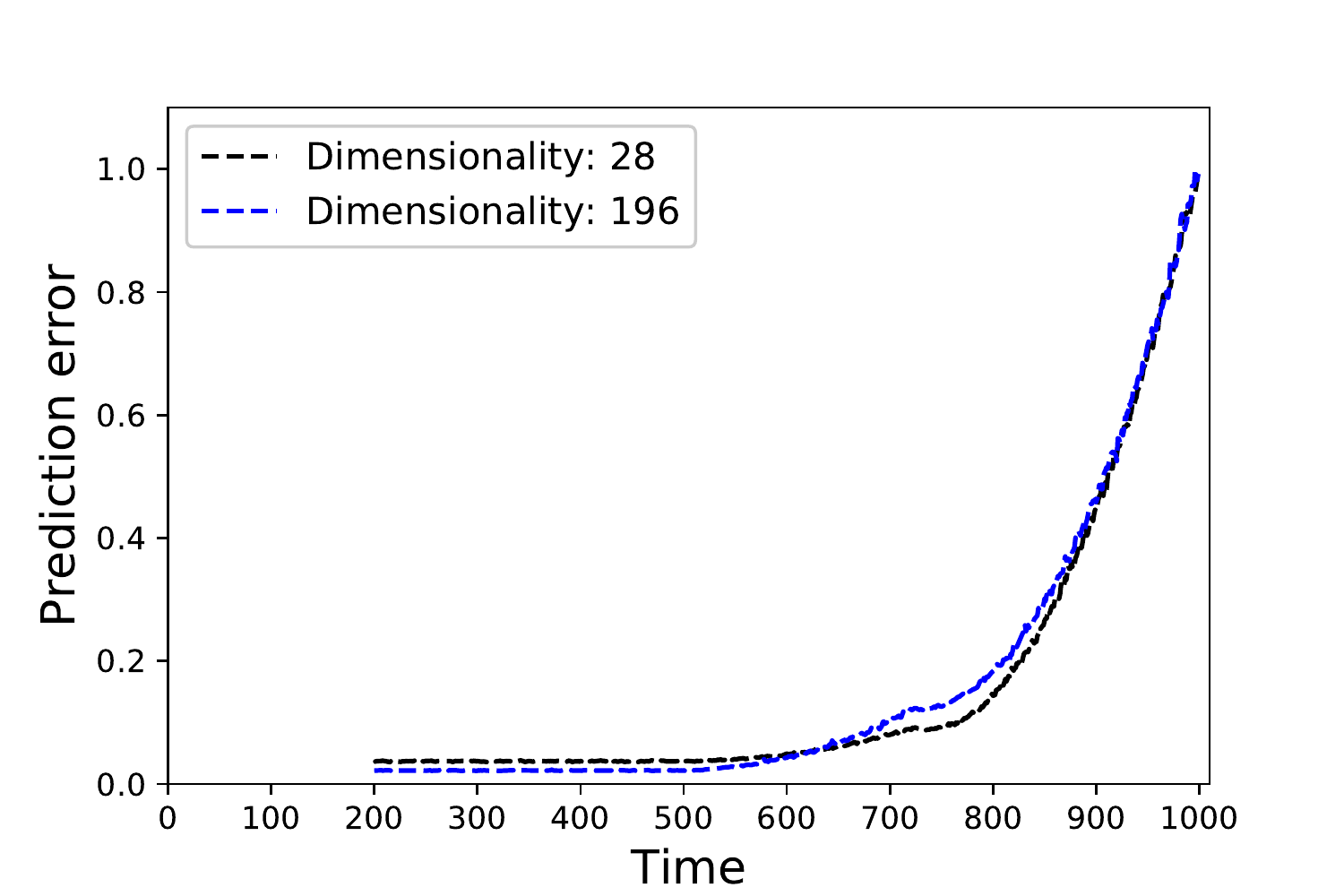}
}
\caption{The anomaly detection results of SAE based approach in Case B.}
\label{fig:case4_accuracy}
\end{figure}
Sensitivity Analysis: Figure \ref{fig:case4_accuracy} shows the anomaly detection results of SAE based approach corresponding to different data dimensionality. The results were normalized into $(0,1]$. It can be observed that, from $t_s=540\sim 550$, the prediction error curves begin to increase gradually, which indicates the anomaly is detected. Meanwhile, the blue dashed line corresponding to high data dimensionality has a higher variance ratio at the detection points, which validates the proposed increasing data dimensionality algorithm is able to improve the detection sensitivity of ML based approach.

\begin{figure}[!t]
\centerline{
\includegraphics[width=2.5in]{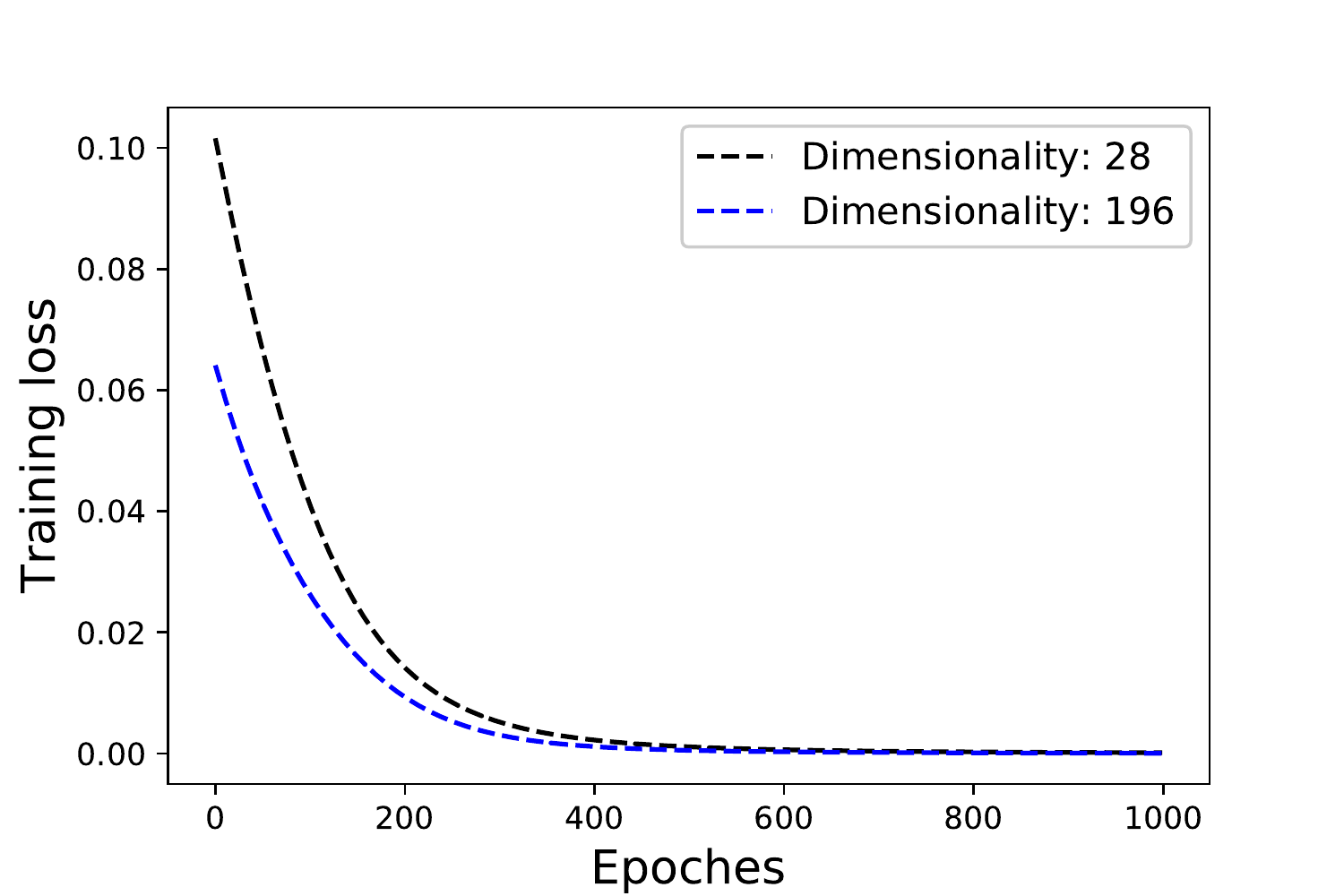}
}
\caption{The convergence rate of SAE based approach in Case B.}
\label{fig:case4_convergence}
\end{figure}
Convergence Analysis: Figure \ref{fig:case4_convergence} illustrates the convergence rate in training SAE for different data dimensionality. It can be observed that the blue dashed line corresponding to high data dimensionality has a faster convergence rate, which validates the proposed increasing data dimensionality algorithm can help improve the convergence rate of model training in SAE based anomaly detection approach.
\subsection{Discussion}
\label{subsection: case_C}
In the proposed data dimensionality increment algorithm in Section \ref{subsection: anomaly_detection}, each column vector of the formulated spatio-temporal data matrix is partitioned into multiple vectors. In theory, those partitioned vectors are required to be independent identically distributed (i.i.d.) for the RMT based anomaly detection approach when a system operates in steady state. However, in practice, the PMUs are installed at different locations in a given power system and the synchrophasors are different each other, which makes it almost impossible for the partitioned vectors to be i.i.d.. In view of this, a residual matrix is often obtained firstly. For example, for the formulated data matrix $\bf D$ in Section \ref{subsection: formulation}, the corresponding residual matrix ${\bf D}'$ is constructed as
\begin{equation}
\label{Eq:residual_matrix}
\begin{aligned}
  d_{i,j}^{'}=d_{i,j+1}-d_{i,j}
\end{aligned},
\end{equation}
where $d_{i,j}$ ($i=1,\cdots,P;j=1,\cdots,N$) are the entries of $\bf D$ and $d_{i,j}^{'}$ are the elements of the constructed residual matrix ${\bf D}'$. In general, $d_{i,j}^{'}$ are considered to be random and approximately Gaussian distributed in steady system state, which can satisfy the i.i.d. prerequisites most for using the RMT based anomaly detection approach.

From Section \ref{subsection: case_A}, it can be concluded that the ML based anomaly detection approach is more powerful in open circuit detection (which is categorized into change point detection by us), because the ML approach is able to automatically learn the complex mapping between the data and their corresponding labels and it does not make any assumptions on the data itself. From Section \ref{subsection: case_B}, it can be concluded the RMT-based anomaly detection approach is capable of detecting the overload anomaly (which is categorized into early anomaly detection by us) in an earlier phase than the ML based approach. The reason is that, for each sampling time, a moving data window instead of only the current sampling data is analyzed in the RMT based approach. The average result makes it more robust against random fluctuations and measuring errors.

\section{Conclusion}
\label{section: conclusion}
Based on the random tensor theory, a data dimensionality increment algorithm is proposed for anomaly detection in low observability power systems. In the RMT based anomaly detection approach, the tensor version LES is introduced and used as the anomaly indicator to indicate the system behavior. In the machine learning based anomaly detection approach, one-class prediction model is trained and used for detecting the anomalies, in which the prediction error is used as the anomaly indicator to indicate the data behavior. The proposed increasing data dimensionality algorithm can help improve the detection sensitivity of RMT based and ML based anomaly detection approaches, and it can accelerate the convergence rate of model training in the ML based anomaly detection approach. Case studies on the IEEE 118-bus test system corroborate the effectiveness of the proposed algorithm, which indicates it can be served as a primitive for PMU data preprocessing in low observability power systems.

\small{}
\bibliographystyle{IEEEtran}
\bibliography{helx}

\normalsize{}
\end{document}